\documentclass[prd,preprint,aps,showpacs]{revtex4}
\usepackage{graphics}
\newcommand{\newc}{\newcommand}
\newc{\gsim}{\lower.7ex\hbox{$\;\stackrel{\textstyle>}{\sim}\;$}}
\newc{\lsim}{\lower.7ex\hbox{$\;\stackrel{\textstyle<}{\sim}\;$}}
\newc{\gev}{\,{\rm GeV}}
\newc{\mev}{\,{\rm MeV}}
\newc{\ev}{\,{\rm eV}}

\newc{\kev}{\,{\rm keV}}
\newc{\tev}{\,{\rm TeV}}
\newc{\mz}{m_Z}
\newc{\mpl}{M_{Pl}}
\newc{\chifc}{\chi_{{}_{\!F\!C}}}
\newc\order{{\cal O}}
\newc\CO{\order}
\newc\CL{{\cal L}}
\newc\CY{{\cal Y}}
\newc\CH{{\cal H}}
\newc\CM{{\cal M}}
\newc\CF{{\cal F}}
\newc\CD{{\cal D}}
\newc\CN{{\cal N}}
\newc{\eps}{\epsilon}
\newc{\re}{\mbox{Re}\,}
\newc{\im}{\mbox{Im}\,}
\newc{\invpb}{\,\mbox{pb}^{-1}}
\newc{\invfb}{\,\mbox{fb}^{-1}}
\newc{\yddiag}{{\bf D}}
\newc{\yddiagd}{{\bf D^\dagger}}
\newc{\yudiag}{{\bf U}}
\newc{\yudiagd}{{\bf U^\dagger}}
\newc{\yd}{{\bf Y_D}}
\newc{\ydd}{{\bf Y_D^\dagger}}
\newc{\yu}{{\bf Y_U}}
\newc{\yud}{{\bf Y_U^\dagger}}
\newc{\ckm}{{\bf V}}
\newc{\ckmd}{{\bf V^\dagger}}
\newc{\ckmz}{{\bf V^0}}
\newc{\ckmzd}{{\bf V^{0\dagger}}}
\newc{\X}{{\bf X}}
\newc{\bbbar}{B^0-\bar B^0}

\newc{\sgn}{\mbox{sgn}\,}
\newc{\m}{{\bf m}}
\newc{\msusy}{M_{\rm SUSY}}
\newc{\munif}{M_{\rm unif}}
\newc{\slepton}{{\tilde\ell}}
\newc{\Slepton}{{\tilde L}}
\newc{\sneutrino}{{\tilde\nu}}
\newc{\selectron}{{\tilde e}}
\newc{\stau}{{\tilde\tau}}
\relax
\def\beq{\begin{equation}}
\def\eeq{\end{equation}}
\def\bea{\begin{eqnarray}}
\def\eea{\end{eqnarray}}
\newc{\ie}{{\it i.e.}}          \newc{\etal}{{\it et al.}}
\newc{\eg}{{\it e.g.}}          \newc{\etc}{{\it etc.}}
\newc{\cf}{{\it c.f.}}
\def\Dsl{\,\raise.15ex\hbox{/}\mkern-13.5mu D} 
\def\delsl{\raise.15ex\hbox{/}\kern-.57em\partial}
\def\Ksl{\hbox{/\kern-.6000em\rm K}}
\def\Asl{\hbox{/\kern-.6500em \rm A}}
\def\Qsl{\hbox{/\kern-.6000em\rm Q}}
\def\gradsl{\hbox{/\kern-.6500em$\nabla$}}
\def\bar#1{\overline{#1}}

\begin{document}
\title{Electroweak-Higgs Unification and the Higgs Boson Mass}
\author{ A. Aranda$^{(a)}$, J.L. D\'{\i}az-Cruz$^{(b)}$, and A. Rosado$^{(c)}$}
\affiliation{$^{(a)}$ Facultad de Ciencias, U. de Colima.\\
Colima, Colima, M\'exico. C.P. 28040 \\
$^{(b)}$ Facultad de Ciencias F\'{\i}sico-Matem\'aticas, BUAP.\\
Apdo. Postal 1364, C.P. 72000 Puebla, Pue., M\'exico\\
$^{(c)}$ Instituto de F\'{\i}sica, BUAP. Apdo. Postal J-48, C.P. 72570 Puebla, Pue., M\'exico}
\date{\today}

\begin{abstract}
We propose  an alternative unification scenario where the Higgs
self-coupling ($\lambda$) is determined by impossing its
unification with the electroweak gauge couplings. An attractive
feature of models within this scenario is the possibility to
determine the Higgs boson mass by evolving $\lambda$ from the
electroweak-Higgs unification scale $M_{GH}$ down to the
electroweak scale. The unification condition for the gauge
($g_1,g_2$) and Higgs couplings is written as
$g_1=g_2=f(\lambda)$, where $g_1=k_Y^{1/2} g_Y$, $k_Y$ being the
normalization constant. Two variants for the unification condition
are discussed; scenario I is defined through the linear relation:
$g_1=g_2=k_H\lambda(M_{GH})$, while scenario II assumes a
quadratic relation: $g^2_1=g^2_2=k_H\lambda(M_{GH})$. Fixing
$k_H=O(1)$ and the standard normalization ($k_Y=5/3$), we obtain a
Higgs boson mass value $m_H \simeq 190$ GeV, with similar results
for other normalizations such as $k_Y=7/4$ and $3/2$. However, the
unification scale $M_{GH}$ depends on the value of $k_Y$, going
from $1.8 \times 10^{12}$ GeV up to $4.9 \times 10^{14}$ GeV for
$7/4 > k_Y > 3/2$. Possible tests of this idea at a future Linear
collider and its application for determining the Higgs spectrum in
the Two-Higgs doublet model are also discussed. We also elaborate
on these unification scenarios within the context of a
six-dimensional $SU(3)_c\times SU(3)_w$ Gauge-Higgs unified model,
where the Higgs boson arises as the extra-dimensional components
of the 6D gauge fields.
\end{abstract}
\pacs{12.60.Fr, 12.15.Mm, 14.80.Cp}

\maketitle

\setcounter{footnote}{0}
\setcounter{page}{2}
\setcounter{section}{0}
\setcounter{subsection}{0}
\setcounter{subsubsection}{0}


\section{Introduction}

The Standard Model (SM) of the strong and electroweak (EW)
interactions has met with extraordinary success; it has been tested
already at the level of quantum corrections
\cite{radcorrs1,radcorrs2}. These corrections give some hints about
the nature of the Higgs sector, pointing towards the existence of a
relatively light Higgs boson, with a mass of the order of the EW
scale,  $m_{\phi_{SM}} \simeq v$ \cite{hixjenser}.
 However, it is widely believed that the SM can not be the final theory
of particle physics,
in particular because the Higgs sector suffers from
naturalness problems, and we really do not have a clear understanding
of electroweak symmetry breaking (EWSB).

These problems in the Higgs sector can be stated as
our present inability to find a satisfactory answer to some
questions regarding its structure. In terms of the
Higgs potential,
\begin{equation}
V(\Phi)= \mu^2_0 \Phi^\dagger \Phi + \frac{\lambda}{4}  (\Phi^\dagger \Phi)^2
\end{equation}
these questions can be stated as follows:
\begin{enumerate}
\item What fixes  the size (and sign) of the  {\it dimensionful parameter}
$\mu^2_0$?. This parameter determines the scale of
EWSB in the SM; in principle it could be as high as the Planck mass,
however it needs to be fixed to much lower values.
\item What is the nature of  the quartic Higgs coupling $\lambda$?.
This parameter is not associated with a known symmetry, and we expect
all interactions in nature to be associated somehow with gauge forces,
as these are the ones we understand better \cite{myghyunif}.
\end{enumerate}

An improvement on our understanding of EWSB is provided by the supersymmetric
(SUSY) extensions of the SM \cite{softsusy}, where loop corrections to the
tree-level parameter  $\mu^2_0$ are under control, thus making the
Higgs sector more natural. The quartic Higgs couplings is nicely
related with gauge couplings through relations of the form:
$\lambda=\frac{1}{8} (g^2_2+g^2_Y)$, thus solving this aspect of the
problem too.
In the SUSY alternative it is even possible to (indirectly) explain
the sign of  $\mu^2_0$ as a result loop effects and
the breaking of the symmetry between bosons and fermions.
Even if the Higgs mass parameter is taken to be positive
at a high energy scale, renormalization effects
drive its value negative at the EW scale, thus
generating (explaining) EWSB.

Further progress to understand the SM structure is achieved in Grand
Unified Theories (GUT), where the strong and electroweak gauge
interactions are unified at a very high-energy  scale ($M_{GUT}$)
\cite{gutrev}. However, certain consequences of the GUT idea seem to
indicate that this  unification, by itself, may be too violent
(within the minimal SU(5) GUT model one actually gets inexact
unification, too large proton decay, doublet-triplet problem, bad
fermion mass relations, etc.), and some additional theoretical tool
is needed to overcome these difficulties. Again, SUSY offers an
amelioration of these problems. When SUSY is combined with the GUT
program, one gets a more precise gauge coupling unification and some
aspects of proton decay and fermion masses are under better control
\cite{myradferm1,myradferm2}.

However, in order to verify the realization of SUSY-GUT in nature,
it will take the observation of plenty of new phenomena such as
superpartners, proton decay or rare decay modes. As nice as these
ideas may appear, it seems worthwhile  to consider other approaches
for physics beyond the SM. For instance, it has been shown that
additional progress towards understanding the SM origin, can be
achieved by postulating the existence of extra dimensions. These
theories have received much attention, mainly because of the
possibility they offer to address the problems of the SM from a new
geometrical perspective. These range from a new approach to the
hierarchy problem \cite{ADD1,ADD2,ADD3,ADD4,RanSundrum} up to a
possible explanation of flavor hierarchies in terms of field
localization along the extra dimensions \cite{nimashmaltz}. Model
with extra dimensions have been applied to neutrino physics
\cite{abdeletal1,abdeletal2,abdeletal3,abdeletal4}, Higgs
phenomenology \cite{myhixXD1,myhixXD2}, among many others. In the
particular GUT context, it has been shown that it is possible to
find viable solutions to the doublet-triplet problem
\cite{hallnomu1,hallnomu2}. More recently, new methods in strong
interactions have also been used as an attempt to revive the old
models (TC, ETC, topcolor, etc) \cite{fathix}. Other ideas have
motivated new types of models as well (little Higgs
\cite{littlehix}, AdS/CFT composite Higgs models \cite{hixdual},
etc).

In  this paper we are interested in exploring alternative
unification scenarios, of weakly-interacting type, which could offer
direct  understanding of the Higgs sector too. We shall explore the
consequences of a scenario where the electroweak $SU(2)_L\times
U(1)_Y$ gauge interactions  are unified with the Higgs
self-interaction at an intermediate scale $M_{GH}$. Namely, we take
seriously the indication of the  running of the SM gauge couplings,
which taken at face value (see Figure 1) show that the $U(1)$ and
$SU(2)$ couplings approach first each other, before crossing with
the strong coupling constant. Thus, we propose two variants for the
unification condition which permit to predict the SM Higgs boson
mass, with resulting values of the order $m_H\simeq 190$ GeV. The
dependence of our results on the choice for the normalization for
the hypercharge is also discussed, as well as possible test of this
EW-Higgs unification idea at future colliders, such as ILC. We
consider then  (section III) the implementation of this idea in the
two-Higgs doublet model, which allows to predict the Higgs spectrum,
namely the masses for the neutral CP-even states ($h^0,H^0$), the
neutral CP-odd state ($A^0$) and the Charged Higgs ($H^\pm$). At
this point it is relevant to compare our approach with the so called
Gauge-Higgs unification program as they share some similarities. We
think that our approach is more model independent, as we first
explore the consequences of a parametric unification, without really
choosing a definite model at higher energies. In fact, at higher
energies both the SUSY models as well as the framework of extra
dimensions could work as ultraviolet completion of our approach. The
SUSY models could work because they allow to relate the scalar
quartic couplings to the gauge couplings, thanks to the D-terms
\cite{myghyunif}. On the other hand, within the extra-dimensions it
is also possible to obtain similar relations, when the Higgs fields
are identified as the extra-dimensional components of gauge fields
\cite{XDGHix1,XDGHix2,ABQuiros1,ABQuiros2,ABQuiros3,Hosotani1,Hosotani2,
Hosotani3,Hosotani4,Hosotani5,Hosotani6,Hosotani7,Hosotani8}.
Actually, we feel that the work of
Ref.\cite{gauhixyuku1,gauhixyuku2} has a similar spirit to ours, in
their case they look for gauge unification of the Higgs
self-couplings that appear in the superpotential of NMSSM, and then
they justify their work with a concrete model in 7D. Thus, in our
case we also discuss the unification of the EW-Higgs couplings with
the strong constant, which can be realized within the context of
extra-dimensional Gauge-Higgs unified theories. In particular we
discuss a 6D $SU(3)_c\times SU(3)_{w}$ model, which is broken on an
orbifold $T^2/Z_3$ and the SM Higgs doublet is identified with the
extra-dimensional components of the gauge fields.

\begin{figure}[floatfix]
\begin{center}
\includegraphics{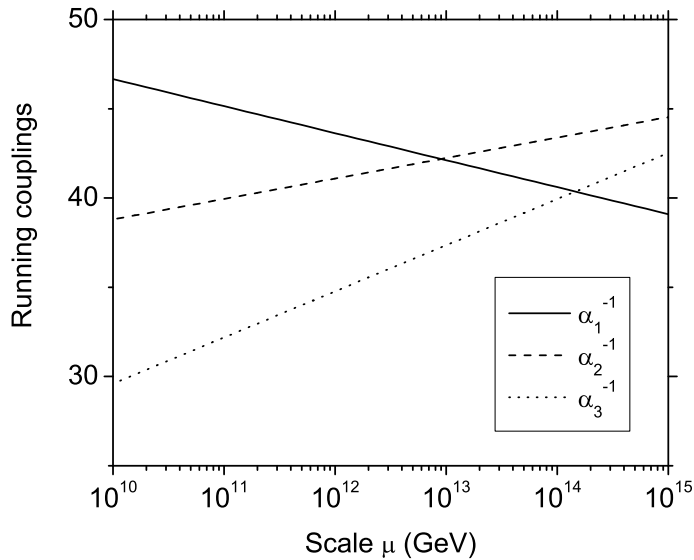}
\caption{Evolution of the running couplings as functions of the
scale $\mu$ in the context of the SM, taking $m_{top}=170.0$ GeV.}
\label{figure1}
\end{center}
\end{figure}

\begin{figure}[floatfix]
\begin{center}
\includegraphics{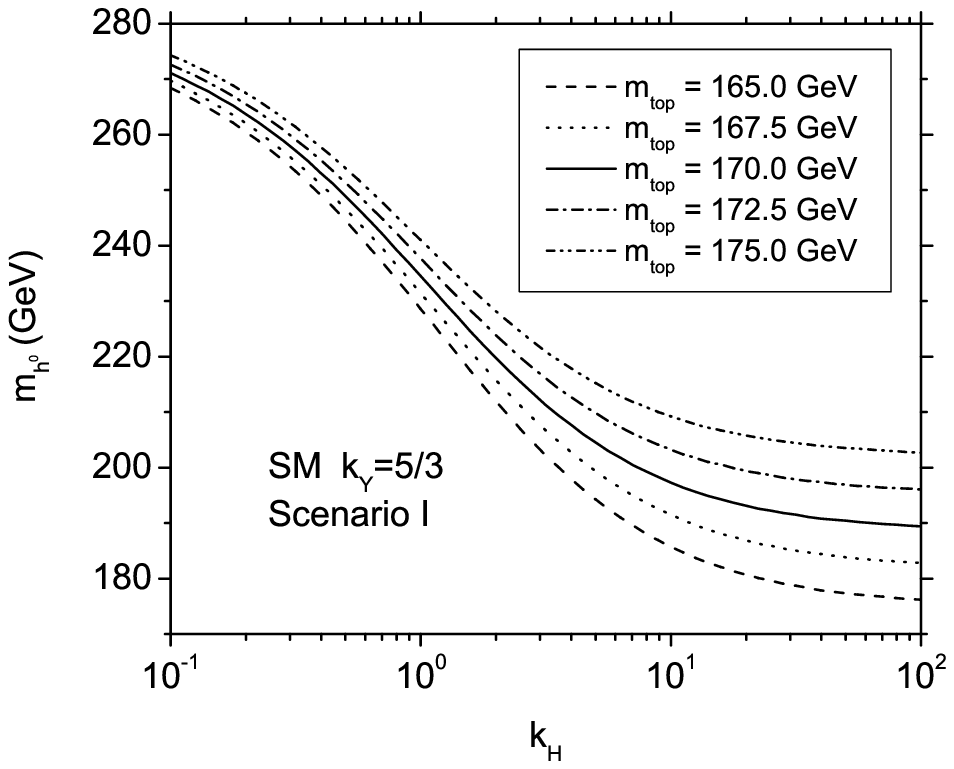}
\caption{Prediction for the Higgs boson mass as a function of
$k_H$ in the context of the SM with $k_Y=5/3$, in the frame of
scenario I, taking $m_{top}=165.0; \, 167.5; \, 170.0; \, 172.5;
\, 175.0$ GeV.} \label{figure2}
\end{center}
\end{figure}

\begin{figure}[floatfix]
\begin{center}
\includegraphics{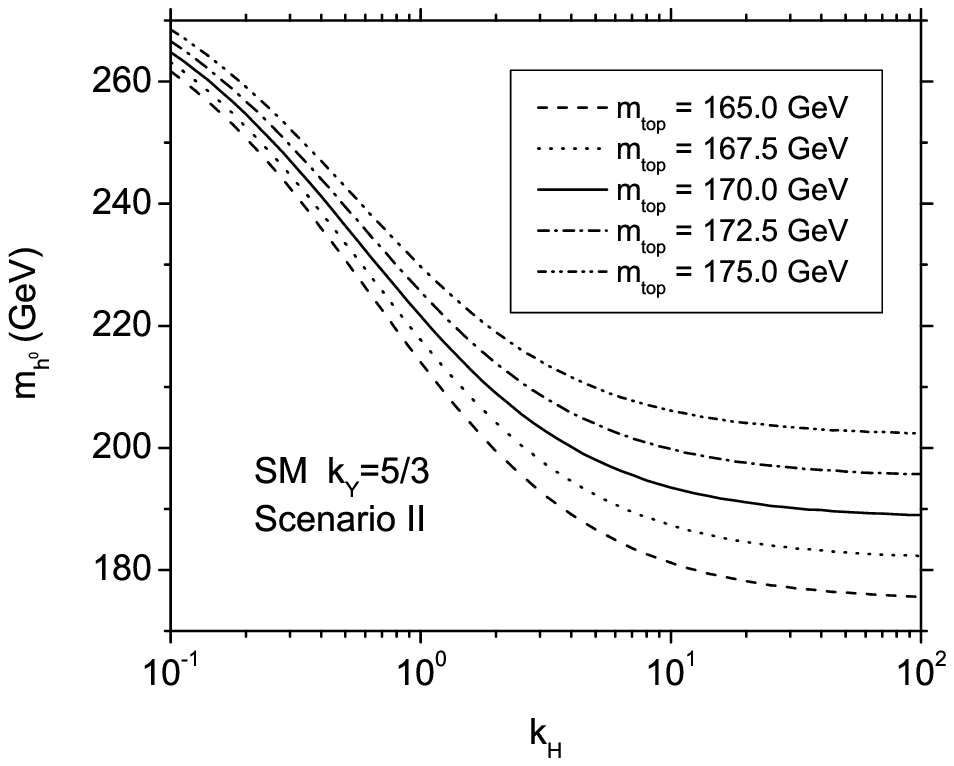}
\caption{Prediction for the Higgs boson mass as a function of
$k_H$ in the context of the SM with $k_Y=5/3$, in the frame of
scenario II, taking $m_{top}=165.0; \, 167.5; \, 170.0; \, 172.5;
\, 175.0$ GeV.} \label{figure3}
\end{center}
\end{figure}

\begin{figure}[floatfix]
\begin{center}
\includegraphics{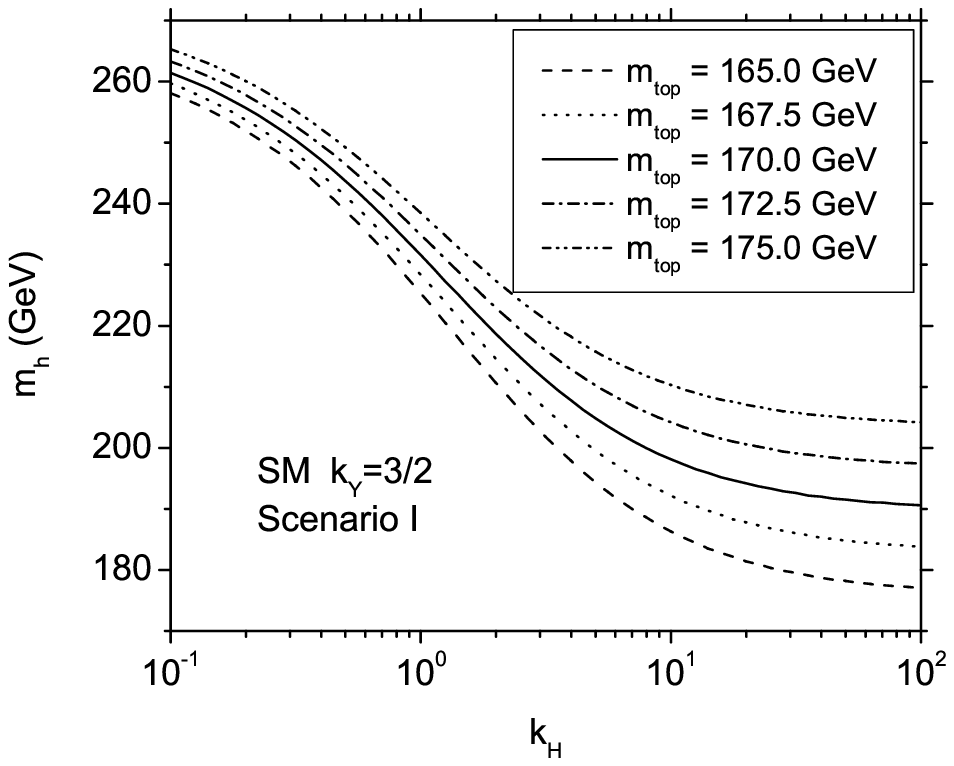}
\caption{Prediction for the Higgs boson mass as a function of
$k_H$ in the context of the SM with $k_Y=3/2$, in the frame of
scenario I, taking $m_{top}=165.0; \, 167.5; \, 170.0; \, 172.5;
\, 175.0$ GeV.} \label{figure4}
\end{center}
\end{figure}

\section{Gauge-Higgs unification in the SM}

In the present EW-Higgs unified scenario, we shall assume that there exists a
scale where the gauge couplings constants $g_1,g_2$, associated with the gauge symmetry
$SU(2)_L\times U(1)_Y$, are unified, and that at this scale they also get unified with
the Higgs self coupling $\lambda$, i.e. $g_1=g_2=f(\lambda)$ at $M_{GH}$.
The precise relation between $g_1$ and $g_Y$ (the SM hypercharge coupling)
involves a normalization factor $k_Y$, i.e. $g_1=k_Y^{1/2}g_Y $, which
depends on the unification model. The standard normalization gives
$k_Y=5/3$, which is associated with minimal models such as $SU(5),
SO(10), E_6$. However, in the context of string theory it is possible to
have such standard normalization without even having a unification
group. For other unification groups that involve additional $U(1)$ factors,
one would have  exotic normalizations too, and similarly for the case
of GUT models in extra-dimensions. In what follows we shall present results
for the cases: $k_Y=5/3,3/2$ and $7/4$, which indeed arise in string-inspired
models \cite{Dienes:1996yh}. Note that these values fall in the range
$3/2 < 5/3 < 7/4$ and so they can
illustrate what happens when one chooses a value below or above the
standard normalization.
The form of the unification condition will depend on the particular
realization of this scenario, which could be as generic as possible.
 However, in order to be able to make predictions for
the Higgs boson mass, we shall consider two specific realizations.
Scenario I will be based on the linear relation:  $g_2=g_1=k_H \lambda(M_{GH})$,
where the factor $k_H$ is included in order to retain some generality, for
instance to take into account possible unknown group theoretical or
normalization factors. Motivated by specific models, such as SUSY
itself, as well as an argument based on the power counting of the
beta coefficients in the RGE for scalar couplings, namely the fact
that $\beta_\lambda$ goes as $O(g^4)$, we shall also define scenario II,
through the quadratic unification condition:
$g^2_1=g^2_2=k_H\lambda (M_{GH})$.

The SM renormalization group equations at the one loop level involving
the gauge coupling constants $g_{1,2,3}$, the Higgs self-coupling
$\lambda$, the top-quark Yukawa coupling $g_t$, and the parameter $k_Y$,
can be written as follows \cite{rgebsm,langa1}:

\begin{eqnarray} \label{2loop1}
 \frac{dg_i}{dt}  &=&  \frac{b_i^{sm}}{16 \pi^2 } g^3_i \ , \\ \nonumber
\\ \label{2loop2}
 \frac{dg_t}{dt}  &=&  \frac{g_t}{16 \pi^2 } \left[ \frac{9}{2} g^2_t -
 \left(\frac{17}{12k_{Y}}g^2_1+ \frac{9}{4}g^2_2+8g^2_3\right)\right] \ , \\
\nonumber \\ \label{2loop3}
 \frac{d\lambda}{dt}  &=&  \frac{1}{16 \pi^2 }\left[12\lambda^2 -
   \left(\frac{3}{k_{Y}}g^2_1+9g^2_2\right)\lambda
+ \frac{9}{4}\left(\frac{1}{3k_{Y}^2}g^4_1+\frac{2}{5}g^2_1g^2_2  +
g^4_2\right) + 12g^2_t\lambda-12g^4_t\right] \ ,
\end{eqnarray}
where $(b_1^{sm},b_2^{sm},b_3^{sm})=(41/6k_{Y},-19/6,-7)$; $\mu$
denotes the scale at which the coupling constants are defined, and
$t=\log(\mu/\mu_0)$. The corresponding expressions for the SM
renormalization group equations at the two-loop level can be found
in Ref.~\cite{langa1}.

Now, we discuss the predictions for the Higgs boson mass and its
dependence on the parameter $k_H$. In practice, we determine first
the scale $M_{GH}$ at which $g_2$ and $g_1$ are unified, then we fix
the quartic Higgs coupling $\lambda$ by imposing the unification
condition and finally, by evolving the quartic Higgs coupling down
to the EW scale, we are able to predict the Higgs boson  mass. For
the numerical calculations, to be discussed next, we employ the full
two-loop SM renormalization group equations involving the gauge
coupling constants $g_{1,2,3}$, the Higgs self-coupling $\lambda$,
the top-quark Yukawa coupling $g_t$, and the parameter $k_Y$
\cite{rgebsm,langa1}. We also take the values for the coupling
constants as reported in the Review of Particle Properties
\cite{revPP}, while for the top quark mass we take the value
recently reported in~\cite{lasttopm1,lasttopm2}.

In order to get a flavor of the coupling constants behavior and evolution,
we find it convenient to discuss first an analytical solution based on the
one-loop beta-functions for the gauge couplings
(Eqs.~(\ref{2loop1}),(\ref{2loop2})), while for the Higgs self-coupling
(Eq.~(\ref{2loop3})) we will only keep the term proportional to $\lambda$
itself, and work with the canonical normalization $k_Y=5/3$.
In this case, the solution for the gauge coupling constants
takes the form:
\begin{equation}
\frac{1}{g^2_i(\mu)}=\frac{1}{g^2_i(\mu_0)}-2 \frac{b_i}{16 \pi^2 } (t-t_0)
\end{equation}
Then, from the unification condition $g_2(M_{GH})=g_1(M_{GH})$,
we get the value: $M_{GH}=1.086\times 10^{13}$ GeV.
Solving the equation for $\lambda$,
we obtain:
\begin{equation}
\frac{1}{\lambda(\mu)}=\frac{1}{\lambda(M_{GH})}+ \frac{3}{4 \pi^2 }
                          ln(M_{GH}/\mu)
\end{equation}
Then, using the relation
$g_2(M_{GH})=g_1(M_{GH})=k_H \lambda(M_{GH})$ from Scenario I, with $k_H=1$,
we get  the value of $\lambda$ at low energies, i.e. for $\mu=m_Z$,
it is $\lambda(m_Z)=0.39$, which implies a Higgs boson mass
of $m_H=219$ GeV. This value is indeed of the order of the EW scale and
the result looks quite encouraging.

Now, let us discuss first the results of the full numerical analysis
for $k_Y=5/3$, for which we find that $M_{GH} \cong 1.0 \times
10^{13}$ GeV. Figures 2 and 3 show the prediction for the Higgs
boson mass in scenarios I and II, respectively. The Higgs boson mass
is shown as function of the parameter $k_H$ over a range $10^{-1} <
k_H < 10^2$, which covers three orders of magnitude and hopefully
illustrates the generality of our results. We want to point out here
that the expected natural value for $k_H$ is 1.

For such a range of $k_H$, the Higgs boson mass takes the values:
$176 < m_H < 275$ GeV for scenario I, while for $k_H=1$ we obtain a
prediction for the Higgs boson mass: $m_H=229,\,234,\,241$ GeV, for
a top quark mass of $m_{top}=165,\,170,\,175$ GeV
\cite{lasttopm1,lasttopm2}, respectively. On the other hand, for
scenario II, we find that the Higgs boson mass can take the values:
$175 < m_H < 269$ GeV, while for $k_H=1$ we obtain:
$m_H=214,\,222,\,230$ GeV.

Then, when we compare our results with the Higgs boson mass obtained
from EW precision measurements, which imply $m_H \lsim 190$ GeV, we
notice that in order to get compatibility with such value, our model
seems to prefer high values of $k_H$. For instance, by taking the
lowest value that we consider here for the top mass, $m_t=165$ GeV,
and fixing $k=10^2$, we obtain the minimum value for the Higgs boson
mass equal to  $m_H=176$ GeV in scenario I, while scenario II
implies a slightly minimal lower value, namely $m_H=175$ GeV.

In turn, Figures 4 and 5, show the results for the Higgs boson
mass using the value $k_Y=3/2$, for which we find that $M_{GH}=
4.9 \times 10^{14}$ GeV, somehow higher than the previous case,
but for which one still gets a mass gap between $M_{GH}$ and  a
possible $M_{GUT}$. One finds somehow higher values for the Higgs
boson mass, for instance for $k_H=1$, one gets
$m_H=225,\,232,\,238$ ($m_H=212,\,220,\,218$) GeV for scenarios I
(II).

On the other hand, Figures 6 and 7, show the results for the Higgs
boson mass using the value $k_Y=7/4$, for which we find that
$M_{GH}= 1.8 \times 10^{12}$ GeV, which is lower than the one of
the previous cases, and has an even larger mass gap between
$M_{GH}$ and  a possible $M_{GUT}$. One finds somehow higher
values for the Higgs boson mass, for instance for $k_H=1$, one
gets $m_H=230,\,236,\,243$ ($m_H=215,\,223,\,231$) GeV for
scenarios I (II).

At this point, rather than continuing discussions on the precise Higgs boson
mass, we would like to emphasize that our approach based on the EW-Higgs
unification idea is very  successful in giving a Higgs boson
mass that has indeed the correct order of magnitude, and that once
measured at the LHC we will be able to fix the
parameter $k_H$ and find connections with other approaches for
physics beyond the SM, such as the one to be discussed next.

In fact, for the Higgs boson mass range that is predicted in our
approach, it turns out that the Higgs will decay predominantly into
the mode $h\to ZZ$, from which there are good chances to meassure
the Higgs boson mass with a precision of the order $5 \%$
\cite{Duhrssen:2004cv1,Duhrssen:2004cv2}, therefore it will make
possible to bound $k_H$ to within the few percent level. Further
tests of our EW-Higgs unification hypothesis would involve testing
more implications of the quartic Higgs coupling. For instance one
could use the production of Higgs pairs ($e^+ e^- \to \nu \bar{\nu}
h h$) at a future linear collider, such as the ILC. This is just
another example of the complementarity of future  studies at LHC and
ILC.

To end this section, we show the evolution of the running couplings
\begin{equation}
\alpha_i^{-1} = \left( \frac{g_i^2}{4 \pi} \right)^{-1} \, \, \, (i=1,2,3) \, ,
\, \, \, \mbox{and} \, \, \left( \frac{k_H^2 \, \lambda^2}{4 \pi} \right)^{-1} \, ,
\end{equation}
as functions of the scale $\mu$ in the context of the SM for
$k_Y=5/3$, $k_Y=3/2$, and $k_Y=7/4$ in Figures 8, 9, 10,
respectively. We perform the numerical calculations in the frame
of scenario I with $k_H=1$ and taking $m_{top}=170.0$ GeV. In
these Figures the EW couplings $g_{1,2}$ unify at the scale
$M_{GH}$ and by assumption so does the Higgs self-coupling
$\lambda$. We notice that in all these cases, the scale $M_{GH}$
is clearly below the usual GUT Scale, thus signaling the viability
of our assumption of EW-Higgs unification.

\section{EW-Higgs unification in the Two-Higgs doublet model.}

In the two-Higgs doublet model (THDM), one includes two scalar
doublets ($\Phi_1$, $\Phi_2$) in the Higgs sector. The Higgs
potential can be written as follows \cite{thdmky}:
\begin{eqnarray}
V(\Phi_1,\Phi_2) &=& \mu^2_1 \Phi_1^\dagger \Phi_1
+ \mu^2_2 \Phi_2^\dagger\Phi_2
+ \lambda_1 (\Phi_1^\dagger \Phi_1)^2
+ \lambda_2 (\Phi_2^\dagger \Phi_2)^2
+ \lambda_3 (\Phi_1^\dagger \Phi_1) (\Phi_2^\dagger \Phi_2) \nonumber\\
 & & + \lambda_4 (\Phi_1^\dagger \Phi_2) (\Phi_2^\dagger \Phi_1)
+ \frac{1}{2} \lambda_5 [(\Phi_1^\dagger \Phi_2)^2 +
    (\Phi_2^\dagger \Phi_1)^2] \ .
\end{eqnarray}

\noindent One can observe that by absorbing a phase in the definition of
$\Phi_2$, one can make $\lambda_5$ real and negative, which pushes all
potential CP violating effects into the Yukawa sector:

\begin{equation}
\lambda_5 \leq 0.
\end{equation}

\noindent In order to avoid spontaneous breakdown of the electromagnetic
$U(1)$ \cite{loren1}, the vacuum expectation values must have the following
form:

\begin{equation}
\langle \Phi_1 \rangle = \left(
\begin{array}{c}
0 \\
v_1
\end{array}
\right) \,\,\, , \,\,\, \langle \Phi_2 \rangle = \left(
\begin{array}{c}
0 \\
v_2
\end{array}
\right) \,\,\, ,
\end{equation}

\noindent $v_1^2 + v_2^2 \equiv v^2 = (246 GeV)^2$. This configuration is
indeed a minimum of the tree level potential if the following conditions
are satisfied.
\begin{eqnarray}
\lambda_1 &\geq& 0 \ , \nonumber \\
\lambda_2 &\geq& 0 \ , \nonumber \\
\lambda_4 + \lambda_5 &\leq& 0 \ , \nonumber \\
4\lambda_1 \lambda_2 &\geq& (\lambda_3 + \lambda_4 + \lambda_5)^2 \ .
\end{eqnarray}

The scalar spectrum in this model, includes two CP-even states ($h^0,H^0$),
one CP-odd ($A^0$) and two charged Higgs bosons ($H^{\pm}$). The tree
level expressions for the masses and mixing angles are given as follows:

\begin{eqnarray}
\tan\beta &=&  \frac{v_2}{v_1} \ , \\
\sin\alpha &=&  -(\mbox{sgn} \, C) \left[\frac{1}{2}
\frac{\sqrt{(A-B)^2+4C^2}-(B-A)}{\sqrt{(A-B)^2+4C^2}} \right]^{1/2} \ , \\
\cos\alpha &=&  \left[\frac{1}{2}
\frac{\sqrt{(A-B)^2+4C^2}+(B-A)}{\sqrt{(A-B)^2+4C^2}} \right]^{1/2} \ , \\
M^2_{H^{\pm}} &=& - \frac{1}{2} (\lambda_4+\lambda_5) v^2 \ , \\
M^2_{A^0} &=& - \lambda_5 v^2 \ , \\
M^2_{H^0,h^0} &=& \frac{1}{2} \left[A+B {\pm} \sqrt{(A-B)^2+4C^2} \right] \ ,
\end{eqnarray}

\noindent where $A=2 \lambda_1 v^2_1$, $B=2 \lambda_2 v^2_2$,
$C=(\lambda_3 + \lambda_4 + \lambda_5) v_1 v_2$.

\noindent The two Higgs doublet models are described by 7 independent
parameters which can be taken to be $\alpha$, $\beta$, $M_{H^{\pm}}$,
$M_{H^{0}}$, $M_{h^{0}}$, $M_{A^{0}}$, while the top quark mass given by:

\begin{equation}
M_t = g_t v \sin\beta \ .
\end{equation}

\noindent Now, we write the THDM renormalization group equations at
the one loop level involving the gauge coupling constants
$g_{1,2,3}$, the Higgs self-couplings $\lambda_{1,2,3,4,5}$, the
top-quark Yukawa coupling $g_t$, and the parameter $k_Y$, as follows
\cite{thdmky,langa1}:

\begin{eqnarray}
\frac{dg_i}{dt}  &=&  \frac{b_i^{thdm}}{16 \pi^2 } g^3_i \ , \\
\frac{dg_t}{dt}  &=&  \frac{g_t}{16 \pi^2 } \left[ \frac{9}{2} g^2_t -
 \left(\frac{17}{12k_{Y}}g^2_1+ \frac{9}{4}g^2_2+8g^2_3\right)\right] \ , \\
\frac{d\lambda_{1}}{dt}  &=&  \frac{1}{16 \pi^2 }\left[24\lambda_{1}^2 +
2\lambda_{3}^2 + 2\lambda_{3}\lambda_{4} + \lambda_{5}^2 +
\lambda_{4}^2 - 3\lambda_{1}(3g^2_2+\frac{1}{k_{Y}}g^2_1) +
12\lambda_{1}g^2_t \right. \nonumber \\
 & & \left. + \frac{9}{8}g^4_2 + \frac{3}{4k_{Y}}g^2_1g^2_2 +
 \frac{3}{8k^2_{Y}}g^4_1 - 6g^4_t \right] \ , \\
 \frac{d\lambda_{2}}{dt}  &=&  \frac{1}{16 \pi^2 }\left[24\lambda_{2}^2 +
2\lambda_{3}^2 + 2\lambda_{3}\lambda_{4} + \lambda_{5}^2 +
\lambda_{4}^2 - 3\lambda_{2}(3g^2_2+\frac{1}{k_{Y}}g^2_1)
\right. \nonumber \\
 & & \left. + \frac{9}{8}g^4_2 + \frac{3}{4k_{Y}}g^2_1g^2_2 +
 \frac{3}{8k^2_{Y}}g^4_1 \right] \ , \\
\frac{d\lambda_{3}}{dt}  &=&  \frac{1}{16 \pi^2 }\left[
4(\lambda_{1} + \lambda_{2})(3\lambda_{3} + \lambda_{4})
+ 4\lambda^2_{3}+2\lambda^2_{4} + 2\lambda_{5}^2
- 3\lambda_{3}(3g^2_2+\frac{1}{k_{Y}}g^2_1) \right. \nonumber \\
& & \left. + 6\lambda_{3}g^2_t+ \frac{9}{4}g^4_2 - \frac{3}{2k_{Y}}g^2_1g^2_2
+ \frac{3}{4k^2_{Y}}g^4_1 \right] \ , \\
\frac{d\lambda_{4}}{dt}  &=&  \frac{1}{16 \pi^2 }\left[
4\lambda_{4}(\lambda_{1} + \lambda_{2} + 2\lambda_{3} + \lambda_{4})
+ 8\lambda_{5}^2
- 3\lambda_{4}(3g^2_2+\frac{1}{k_{Y}}g^2_1) \right. \nonumber \\
& & \left. + 6\lambda_{4}g^2_t+ \frac{3}{k_{Y}}g^2_1g^2_2
\right] \ , \\
\frac{d\lambda_{5}}{dt}  &=&  \frac{1}{16 \pi^2 }\left[\lambda_{5}
\left((4\lambda_{1} + 4\lambda_{2} + 8\lambda_{3} + 12\lambda_{4}
+ 8\lambda_{5}^2 - 3(3g^2_2+\frac{1}{k_{Y}}g^2_1) + 6g^2_t \right)
\right] \ ,
\end{eqnarray}
where $(b_1^{thdm},b_2^{thdm},b_3^{thdm})=(7/k_{Y},-3,-7)$; $\mu$
denotes the scale at which the coupling constants are defined, and
$t=\log(\mu/\mu_0)$.

\noindent The form of the unification condition will depend on the particular
realization of this scenario, which could be as generic as possible.
However, in order to be able to make predictions for
the Higgs mass, we shall consider again two specific realizations.
Scenario I will be based on the linear relation:
\begin{equation}
g_1=g_2=k_H(i) \, \lambda_i(M_{GH}) \hspace{2.5cm} (i=1,2,3,4,5) \, ,
\end{equation}
where the factors $k_H(i)$ are included in order to take into account
possible unknown group theoretical or normalization factors. We shall
also define scenario II, which uses quadratic unification conditions;
as follows:
\begin{equation}
g^2_1=g^2_2=k_H(i) \, \lambda_i(M_{GH}) \hspace{2.5cm} (i=1,2,3,4,5) \, .
\end{equation}

Now, we present first the results of the numerical analysis for
the Higgs bosons masses in the context of the two Higgs-doublet
model for $\tan\beta=1$ and taking $m_{top}=170.0$ GeV. In order
to get an idea of the behavior of the masses of the Higgs bosons
($h^0$, $H^0$, $A^0$, $H^{\pm}$) we make the following choice {\it
ad hoc}:
\begin{equation}
-k_H(5)=- \frac{1}{3} k_H(4)=3 k_H(3)=k_H(2)=k_H(1) \, ,
\end{equation}
for both scenarios I and II. The Higgs bosons masses are shown as
functions of the parameter $k_H$ over a range $0.1 < k_H < 10$,
which covers two orders of magnitude and hopefully illustrates the
generality of our results.

Now, let us discuss first the results of the full numerical analysis for
$k_Y=5/3$, for which we find that $M_{GH}= 1.3 \times 10^{13}$ GeV.
In Figures 11 and 12 we show the prediction for the Higgs bosons masses
as functions of $k_H(1)$ for $k_Y=5/3$, in the frame of scenarios I and II,
respectively.

For such a range of $k_H(1)$, the mass of the Higgs boson $h^0$
takes the values: $m_{h^0} < 98$ GeV for scenario I, while for
$k_H(1)=1$ we obtain a prediction for the Higgs boson mass:
$m_{h^0}=97$ GeV. On the other hand, for scenario II, we find that
the Higgs boson mass can take the values:  $m_{h^0} < 98$ GeV,
while for $k_H(1)=1$ we obtain: $m_{h^0}=88$ GeV.

In turn, Figures 13 and 14, show the results for the Higgs boson
mass using the value $k_Y=3/2$, for which we find that $M_{GH}= 5.9
\times 10^{14}$ GeV, somehow higher than the previous case, but for
which one still gets a mass gap between $M_{GH}$ and  a possible
$M_{GUT}$. One finds somehow lower values for the Higgs boson mass,
for instance for $k_H(1)=1$, one gets $m_{h^0}= 92$ (85) GeV for
scenarios I (II).

On the other hand, Figures 15 and 16, show the results for the
Higgs boson mass using the value $k_Y=7/4$, for which we find that
$M_{GH}= 2.2 \times 10^{12}$ GeV, which is lower than the one of
the previous cases, and has an even larger mass gap between
$M_{GH}$ and  a possible $M_{GUT}$. One finds somehow higher
values for the Higgs boson mass, for instance for $k_H(1)=1$, one
gets $m_{h^0}= 98$ (88) GeV for scenarios I (II).

As can be seen from these figures, for lower values of $\tan\beta$
one gets an irregular behavior of the other Higgs masses, while for
higher values of $\tan\beta$, there is an approximate degeneracy,
with all masses in the range of $100 - 120$ GeV. Finally, we want to
put emphasis  on the following. Even though the analysis of the
EW-Higgs Unification within the THDM implies that the lightest
neutral CP-even Higgs boson has a mass ($\sim 80$ GeV) that is
somehow below the LEP bounds, 114.4 GeV
\cite{unknown:2006cr,Barate:2003sz}, it should be mentioned that
this bound refers to the SM Higgs boson. The bound on the lightest
Higgs boson of the THDM depends on the factor
$\sin^2(\beta-\alpha)$, which could be less than 1, and thus results
in weaker Higgs boson mass bounds. We have estimated this factor
within our approach and we obtain typical values in the range $0.01
\leq \sin^2(\alpha-\beta) \leq 0.8$, (see Table~\ref{tab:t1}) which
means that those LEP bounds do not necessarily apply. In fact, we
can make use of the experimental results reported in the Table 14 of
Ref.\cite{unknown:2006cr} which allow assuming SM decay rates a
simple and direct check of our model results. After this comparison
we observe that only the first four points in our Table I survive.
Even though that the parameter space is drastically reduced, this
shows that there is still an allowed region which deserves future
detailed investigations. We end this section saying that for the
THDM case our model seems to prefer values of $k_H \lsim 1$.

\section{EW-Higgs and Extended GUT within extra-dimensions.}

We shall show in this section that it is possible to make further
progress: to achieve a Grand Unification of all the coupling
constants at an scale $M_{GUT} > M_{GH}$,
and to justify the EW-Higgs unification. All we need to
do is to consider an extra-dimensional setting.

We have already shown the evolution of the coupling constants in
Figures 8, 9 and 10. The EW couplings $g_{1,2}$ unify at the scale $M_{GH}$
and by assumption so does the Higgs self-coupling $\lambda$.
Above $M_{GH}$ we shall assume that $g_{1,2}$ and $\lambda$ evolve
as a single unified coupling $g_{H}$. However, to determine its
evolution, we need to assume some particular way of embedding the
groups $SU(2)_L$ and $U(1)_Y$ into some unified gauge group, $G_{WH}$.
Although it is possible to construct 4D models where such unification
can be achieved, we also want to justify our assumption of EW-Higgs
unification and, at present, the most promising approach seems to be
the Gauge-Higgs unification in extra-dimensions.
Furthermore, in order to search for a complete unification
of $g_H$ with the strong coupling constant $g_3$, we also need to
assume what kind of physics exists above $M_{GH}$ for the QCD sector.

Here is where extra-dimensions will play a prominent role. We assume
that above $M_{GH}$ the window into a pair of extra-dimensions opens
up. Thus, both the strong and the EW-Higgs couplings will run from
$M_{GH}$ up to another unification scale $M_{GUT}$, with the typical
power-like behavior of extra-dimensional running of the coupling
constants. Furthermore, the extra-dimensional (XD) framework
provides a possible explanation for the EW-Higgs unification.
Namely, we shall identify the Higgs boson as a component of the XD
gauge field \cite{XDGHix1,XDGHix2}. Promising models could be
constructed in five and six dimensions, with or without SUSY
\cite{ABQuiros1,ABQuiros2,ABQuiros3,Hosotani1,Hosotani2,
Hosotani3,Hosotani4,Hosotani5,Hosotani6,Hosotani7,Hosotani8}.
Unification of Higgs and matter with the gauge multiplets has also
been discussed \cite{gauhixyuku1,gauhixyuku2}.

 To build a model based on the idea of gauge-Higgs unification in
extra-dimensions,
we consider a 6D $SU(3)_c\times SU(3)_W$ gauge theory
compactified on a $T^2/Z_n$ orbifold. For $Z_n=Z_2$ one gets two Higgs
doublets in the spectrum of zero modes, while for $Z_n=Z_3$ one gets only
one Higgs doublet. Here we shall consider the case $Z_3$ for definitness.
The 6D gauge bosons are $A_M= T^A A^A_M$  [$M=(\mu,i)$, with $i=5,6$].
The full gauge symmetry is broken by the orbifold
boundary conditions (O.B.C.):
\begin{eqnarray} \nonumber
A_\mu(x_\mu,y) \to A_\mu(x_\mu,-y)&=& +P A_\mu(x_\mu,y)P^{-1}, \\
A_i(x_\mu,y) \to A_i(x_\mu,-y)  &=& - P A_i(x_\mu,y)P^{-1},
\end{eqnarray}
where $P=diag(1,1,-1)$ acts on gauge space as an ``inner automorphism'',
such that the gauge symmetry can be broken as: $G \to H$, with $H=SM$.
Thus, O.B.C. split the group generators into two sets,
$T^A=\{T^a, T^k\}$, $T^A \, \epsilon \, G$, $T^a \, \epsilon \,H$.
Since $A^a_\mu$
has even $Z_2-$ parity, it has zero modes in the spectrum.
On the other hand, $A^k_\mu$ has odd $Z_2-$ parity,
and does not have zero modes in the spectrum.
Furthermore, $A^a_i$  (odd-odd) has zero modes,
and its v.e.v. can break the symmetry further,
namely from $H \to H'=SU(3)_c\times U(1)_{em}$ (for a more
complete discussion of this model, but with a compactification scale
of $O(TeV^{-1}$, see our coming article \cite{futurework}).

Turning now to the gauge coupling unification, we notice that above
$M_{GH}$ the evolution of both the $SU(3)_c$ and $SU(3)_W$ gauge
couplings will need to incorporate the effects of the KK modes. To
describe such effects we shall use the formulas presented
in~\cite{hallnomu1,hallnomu2}, and we shall also incorporate the
fermion content discussed in~\cite{ABQuiros1,ABQuiros2,ABQuiros3},
namely the LH leptons make a triplet under $SU(3)_W$, for which one
needs to add an exotic lepton, while the LH quarks and the RH
up-type quarks will also behave as triplets under $SU(3)_W$, in
addition one needs to include their mirror partners, as it is
required from considering chiral fermions in six dimensions. We
shall also assume that either the inclusion of additional $U(1)$
factor or brane-kinetic terms, make possible to use the canonical
normalization $k_Y=5/3$. Then, the gauge constants
$\alpha_H(\mu)=g^2_{H}/4\pi$ and $\alpha_c(\mu)=g^2_c/4\pi$, at
scales $\mu > M_{GH}$, are related to their values at the scale
$M_{GH}$, through the one-loop expressions:
\begin{eqnarray}
\frac{1}{\alpha_c(\mu)} &=& \frac{1}{\alpha_c(M_{GH})}
-\frac{b_c}{2\pi} ln( \frac{\mu}{M_{GH}}) -
\frac{\tilde{b}_c}{2\pi} F^c_{KK}\, ,\\
\frac{1}{\alpha_w\mu)} &=& \frac{1}{\alpha_w(M_{GH})}
-\frac{b_w}{2\pi} ln( \frac{\mu}{M_{GH}}) - \frac{\tilde{b}_w}{2\pi}
F^w_{KK} \, .
\end{eqnarray}

The terms proportional to $b_i$ represent the effects of the zero
modes, while those proportional to  $\tilde{b}_i$ correspond to the
effects of the KK modes. For the strong group we obtain:
$b_c=-11+4=-7$, the factor -11 comes from pure QCD,
while the factor +4 arises from the 12 Weyl quark fields.
Also  $\tilde{b}_c=-11+1=-10$, and here -11 represents the effects of
the KK modes of the gluon, while +1 comes from the pair of adjoint
colored scalars that come with the massive KK gluons.
For the EW-Higgs sector we obtain: $b_w=-2$, $\tilde{b}_w=-10$;
now $b_w$ is smaller than $b_c$ because in addition to the
quarks we also have to consider the leptons as triplets of $SU(3)_w$,
as well as their mirror partners,
which amount to include a total of 24 Weyl fermion fields.
On the other hand, the function $F^i_{KK}$ represents the sum over the
KK modes, and in the continuous limit it can be approximated as:
$F^i_{KK}=\frac{\pi}{2}(\frac{\mu}{M_{GH}})^2$.

The solution to these equations are matched to the ones for $g_i$ at
$M_{GH}$, and are shown in Figure 19. One can notice that
unification of $g_{GH}$ and $g_3$ appears to occur already at an
scale $M \simeq 10^{14}$, however this is just the scale where the
couplings start to approach each other. In fact, the solution to the
equation $\alpha_H=\alpha_c$, gives a scale $M_{GUT} = 4.5 \times
10^{15}$ GeV (see Figure 19 and Table~\ref{tab:t2}), which we think
represents better the scale at which the group $G=SU(3)_c\times
SU(3)_W$ unifies into some larger group, for instance $SU(6)$, which
may or not induce a large rate for proton decay. In any case, the
GUT scale obtained above is still above the one obtained from
current limits on proton decay induced by dimension-six operators
\cite{gutrev}. We want to point out that according to Eqs. (30) and
(31) the value of $M_{GUT}$ depends only on the value of $M_{GH}$,
if we assume an uncertainty of 10\% (50\%) on the value of $M_{GH}$
i.e. $0.9 \times 10^{13} \, \mbox{GeV} \leq M_{GH} \leq 1.1 \times
10^{13} \, \mbox{GeV}$ ($0.5 \times 10^{13} \, \mbox{GeV} \leq
M_{GH} \leq 1.5 \times 10^{13} \, \mbox{GeV}$) this leads to a
similar uncertainty in $M_{GUT}$, namely we obtain $4.0 \times
10^{15} \, \mbox{GeV} \leq M_{GUT} \leq 5.0 \times 10^{15} \,
\mbox{GeV}$ ($2.2 \times 10^{15} \, \mbox{GeV} \leq M_{GUT} \leq 6.7
\times 10^{15} \, \mbox{GeV}$). We have checked numerically that the
value of $M_{GH}$ does not depend strongly on experimental errors of
the low-energy values of the coupling constants and parametric
errors from the SM input.

We have already pointed out in section II that in order to perform
our numerical calculations we employ the full two-loop SM
renormalization group equations involving the gauge coupling
constants $g_{1,2,3}$, the Higgs self-coupling $\lambda$, the
top-quark Yukawa coupling $g_t$, and the parameter $k_Y$. However,
we have checked that the one-loop SM renormalization group equations
dominate by far the evolution of the $g_{1,2,3}$ $, \lambda$ and
$g_t$ couplings. Therefore, we can expect that the value of $M_{GH}$
and hence the value of $M_{GUT}$ do not depend on missing
higher-order corrections in the renormalization group equations.

On the other hand, as we have already seen $M_{GH}$ depends strongly
on the normalization constant $k_Y$. In fact, for $k_Y=7/4$ (i.e.
$M_{GH} = 1.8 \times 10^{12}$ GeV) we get $M_{GUT} = 7.9 \times
10^{14}$ GeV, and for $k_Y=3/2$ (i.e. $M_{GH} = 4.9 \times 10^{14}$
GeV) we get $M_{GUT} = 2.2 \times 10^{17}$ GeV.

\section{Comments and conclusions}

In this paper we have discussed a framework where it is possible
to unify the Higgs self-coupling with the gauge interactions.
Working first within a phenomenological approach we use this idea
to derive a prediction for the Higgs mass, which is of the order
$m_H\simeq 190$ GeV. Then, we showed how this simple idea
 can be realized within the context of
extra-dimensional theories, where it is also possible to achieve
extended unification at a correct GUT scale.
Thus, we have succeeded in identifying the Higgs self-interaction as
another manifestation of gauge symmetry.

The hypercharge normalization plays an important role to identify
the EW-Higgs unification scale. For the canonical value $k_Y=5/3$ we
get $M_{GH} = 1.0 \times 10^{13}$ GeV. For lower values, such as
$k_Y=3/2$ the scale is $M_{GH} = 4.9 \times 10^{14}$ GeV, which is
closer to the GUT one ($\approx 10^{16}$ GeV) but for higher values,
such as $k_Y=7/4$, which gives $M_{GH} = 1.8 \times 10^{12}$ GeV,
the EW-Higgs unification becomes clearly distinctive.

The present approach still lacks a solution to the hierarchy problem;
at the moment we have to affiliate to argument that fundamental
physics could accept some fine tuning \cite{splitsusy}.
Another option would be to consider one of the simplest early
attempts to solve the problem of quadratic divergences in the SM,
namely through an accidental cancellation~\cite{veltmanqd}. In fact,
such kind of cancellation implies a relationship between the quartic
Higgs coupling and the Yukawa and gauge constants, which has the form:
$\lambda = y^2_t -\frac{1}{8} [ 3g^2 +g'^2]$.
Unfortunately, this relation implies a Higgs
mass $m_\phi=316$ GeV, and that seems already excluded.
Nevertheless, this relation could work if one takes into
account the running of the coupling and Yukawa constants,
a point that we leave for future investigations.

\acknowledgments{The authors would like to thank
CONACYT and SNI (Mexico) for financial support. J.L.  Diaz-Cruz and A. Rosado
thank the Huejotzingo Seminar for inspiring discussions.
J.L. Diaz-Cruz also thanks J. Erler for interesting discussions.}


\newpage

\begin{figure}[floatfix]
\begin{center}
\includegraphics{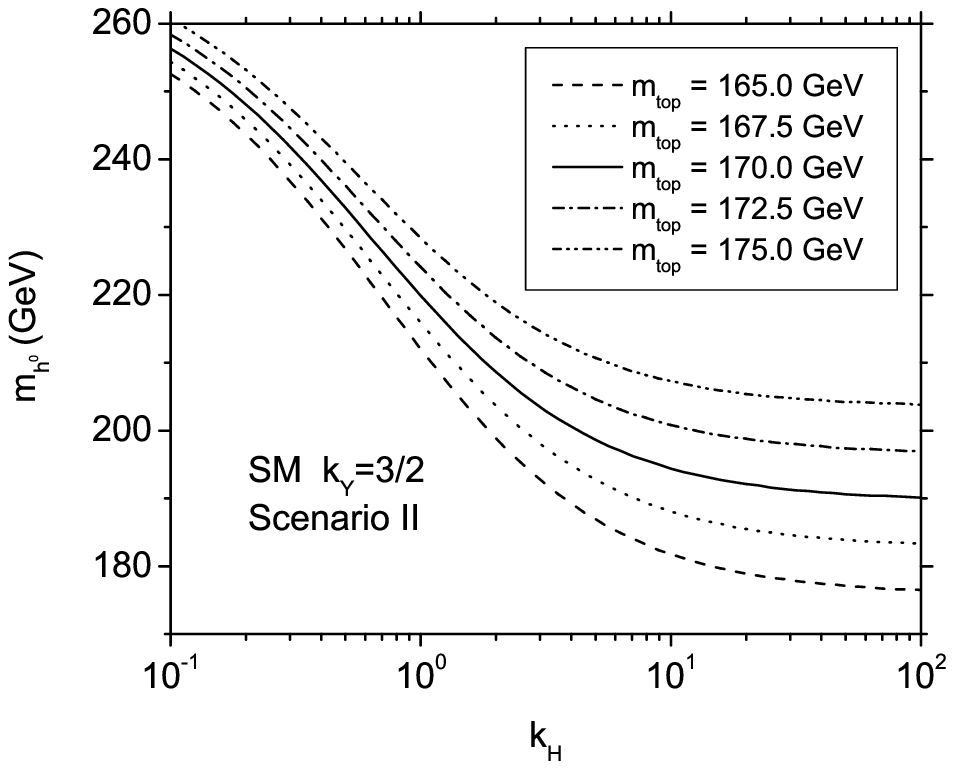}
\caption{Prediction for the Higgs boson mass as a function of
$k_H$ in the context of the SM with $k_Y=3/2$, in the frame of
scenario II, taking $m_{top}=165.0; \, 167.5; \, 170.0; \, 172.5;
\, 175.0$ GeV.} \label{figure5}
\end{center}
\end{figure}

\begin{figure}[floatfix]
\begin{center}
\includegraphics{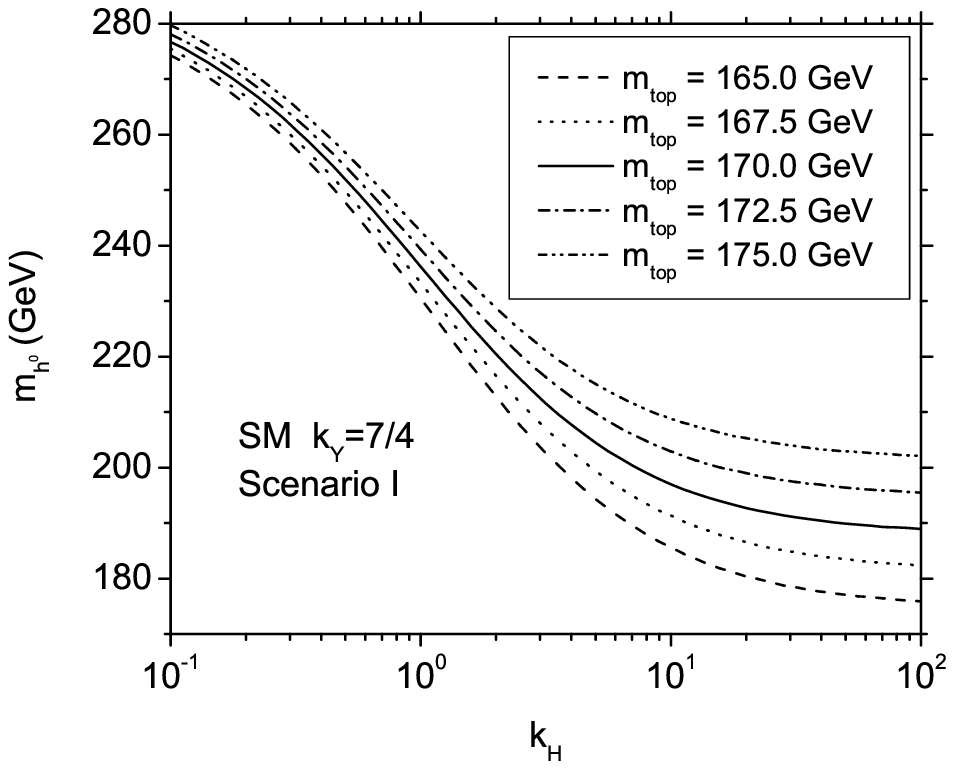}
\caption{Prediction for the Higgs boson mass as a function of
$k_H$ in the context of the SM with $k_Y=7/4$, in the frame of
scenario I, taking $m_{top}=165.0; \, 167.5; \, 170.0; \, 172.5;
\, 175.0$ GeV.} \label{figure6}
\end{center}
\end{figure}

\begin{figure}[floatfix]
\begin{center}
\includegraphics{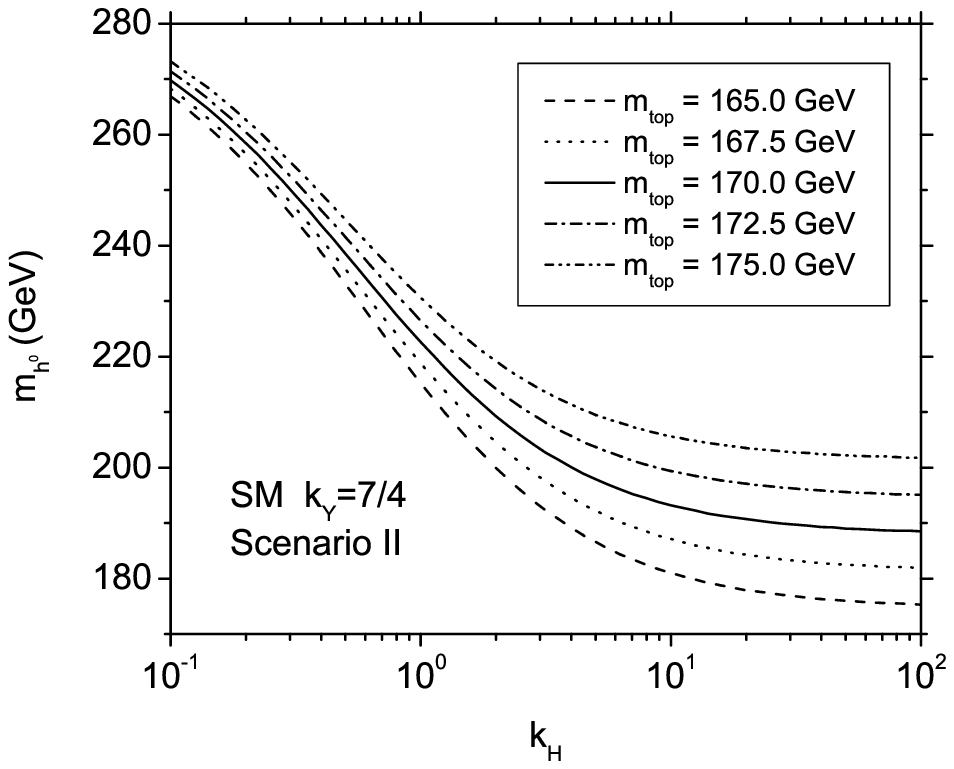}
\caption{Prediction for the Higgs boson mass as a function of
$k_H$ in the context of the SM with $k_Y=7/4$, in the frame of
scenario II, taking $m_{top}=165.0; \, 167.5; \, 170.0; \, 172.5;
\, 175.0$ GeV.} \label{figure7}
\end{center}
\end{figure}

\begin{figure}[floatfix]
\begin{center}
\includegraphics{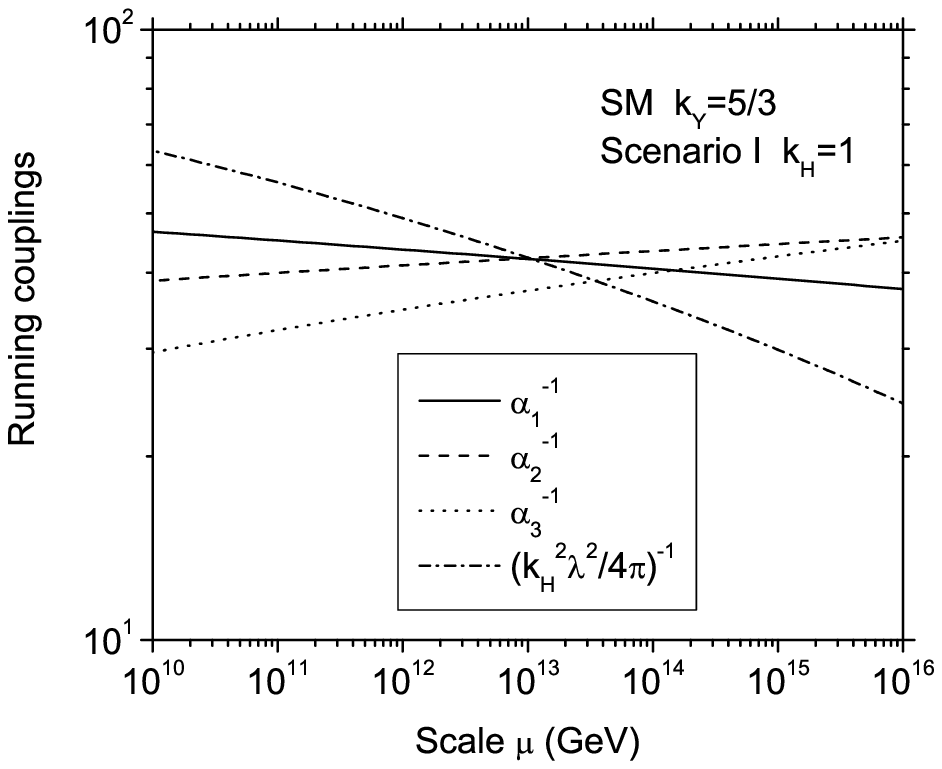}
\caption{Evolution of the running couplings as functions of the
scale $\mu$ in the context of the SM with $k_Y=5/3$, in the frame
of scenario I with $k_H=1$, taking $m_{top}=170.0$ GeV.}
\label{figure8}
\end{center}
\end{figure}

\begin{figure}[floatfix]
\begin{center}
\includegraphics{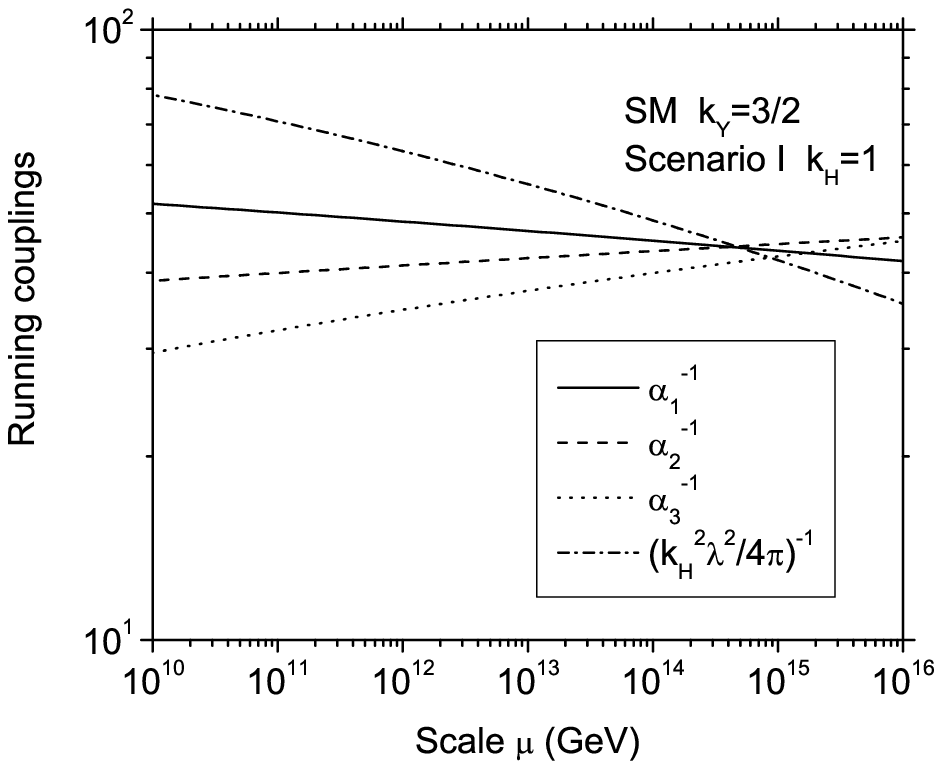}
\caption{Evolution of the running couplings as functions of the
scale $\mu$ in the context of the SM with $k_Y=3/2$, in the frame
of scenario I with $k_H=1$, taking $m_{top}=170.0$ GeV.}
\label{figure9}
\end{center}
\end{figure}

\begin{figure}[floatfix]
\begin{center}
\includegraphics{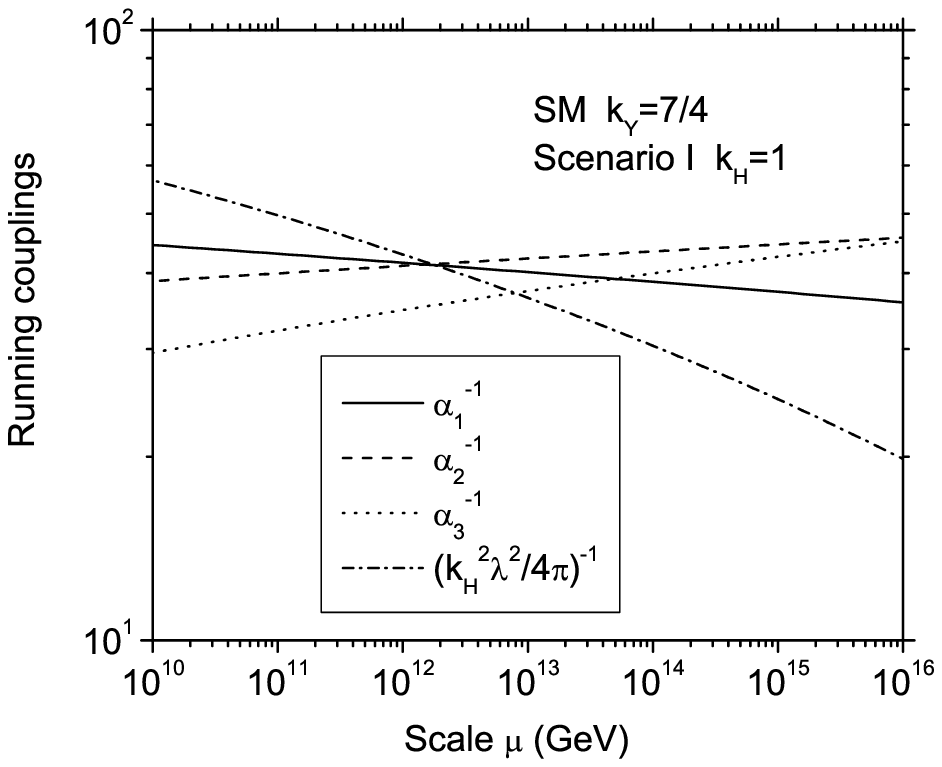}
\caption{Evolution of the running couplings as functions of the
scale $\mu$ in the context of the SM with $k_Y=7/4$, in the frame
of scenario I with $k_H=1$, taking $m_{top}=170.0$ GeV.}
\label{figure10}
\end{center}
\end{figure}

\begin{figure}[floatfix]
\begin{center}
\includegraphics{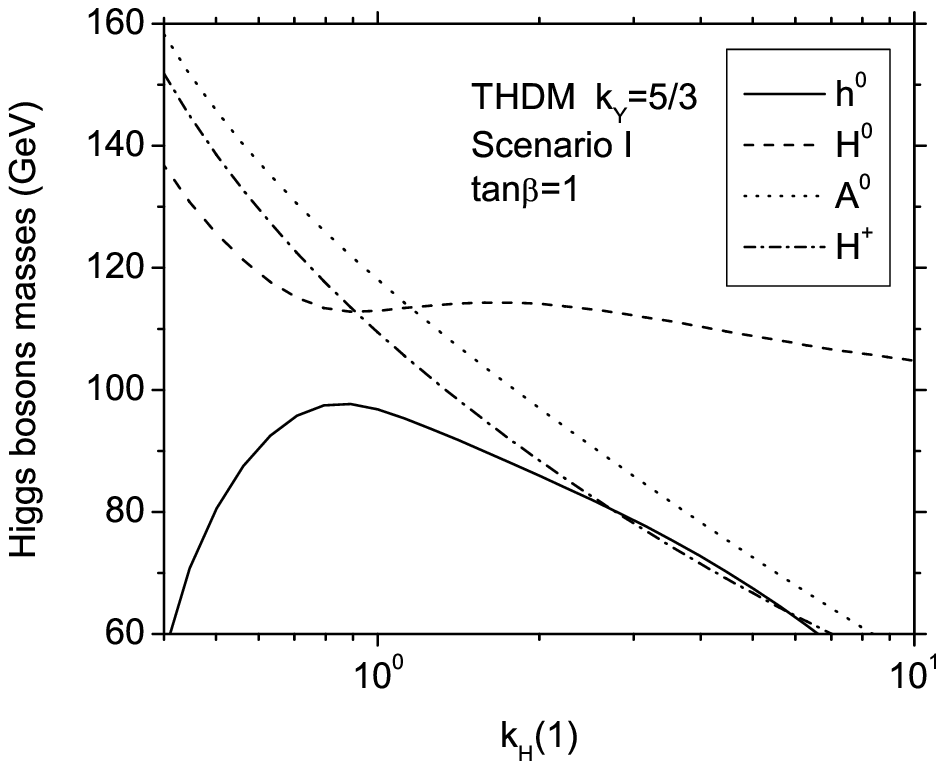}
\caption{Prediction for the Higgs bosons masses as functions of
$k_H(1)$ in the context of the THDM with $k_Y=5/3$, in the frame
of scenario I with $\tan\beta=1$, taking $m_{top}=170.0$ GeV.}
\label{figure11}
\end{center}
\end{figure}

\begin{figure}[floatfix]
\begin{center}
\includegraphics{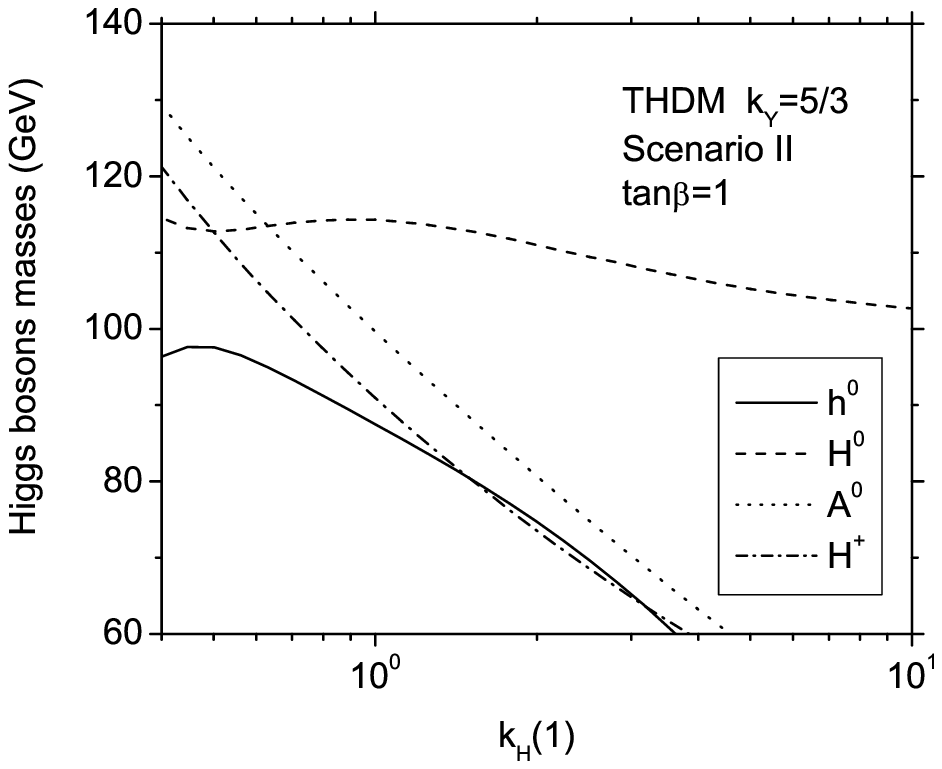}
\caption{Prediction for the Higgs bosons masses as functions of
$k_H(1)$ in the context of the THDM with $k_Y=5/3$, in the frame
of scenario II with $\tan\beta=1$, taking $m_{top}=170.0$ GeV.}
\label{figure12}
\end{center}
\end{figure}

\begin{figure}[floatfix]
\begin{center}
\includegraphics{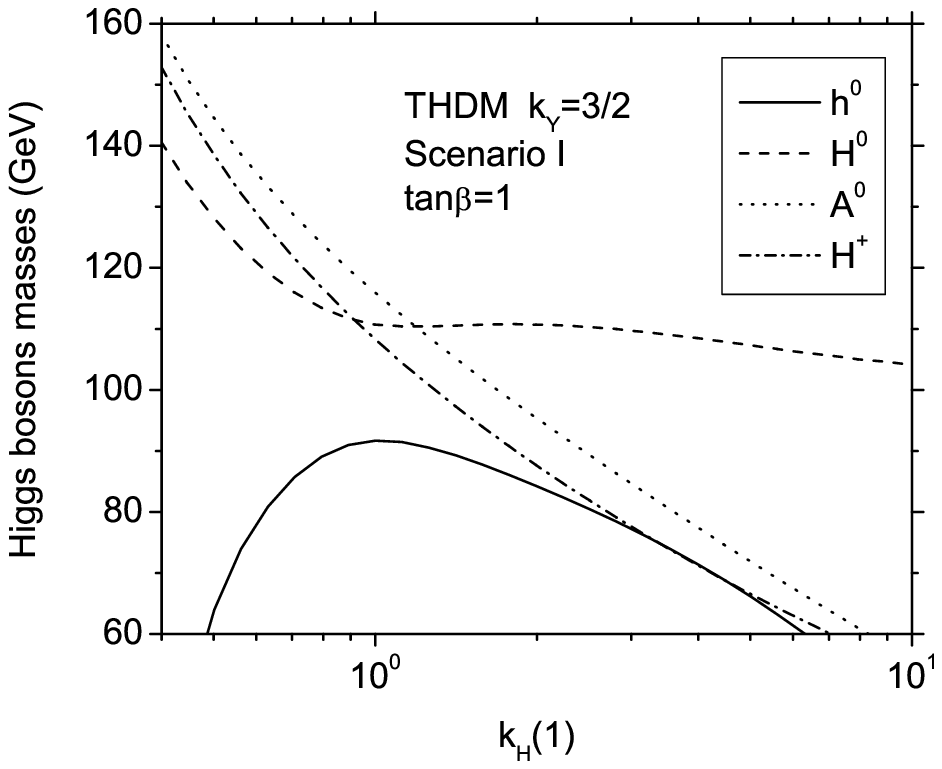}
\caption{Prediction for the Higgs bosons masses as functions of
$k_H(1)$ in the context of the THDM with $k_Y=3/2$, in the frame
of scenario I with $\tan\beta=1$, taking $m_{top}=170.0$ GeV.}
\label{figure13}
\end{center}
\end{figure}

\begin{figure}[floatfix]
\begin{center}
\includegraphics{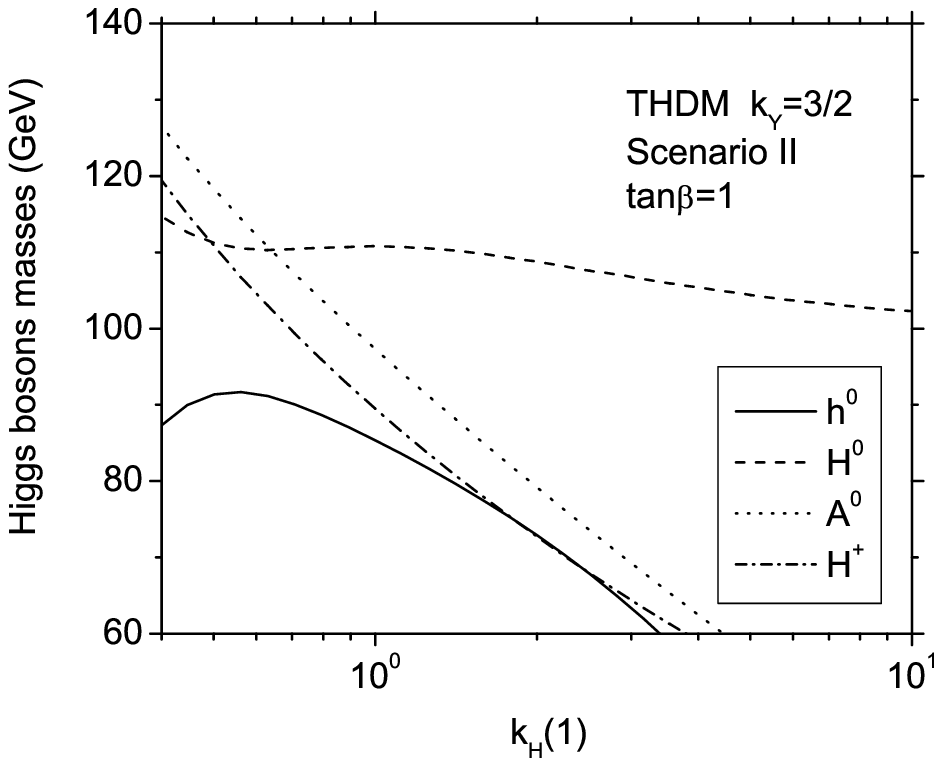}
\caption{Prediction for the Higgs bosons masses as functions of
$k_H(1)$ in the context of the THDM with $k_Y=3/2$, in the frame
of scenario II with $\tan\beta=1$, taking $m_{top}=170.0$ GeV.}
\label{figure14}
\end{center}
\end{figure}

\begin{figure}[floatfix]
\begin{center}
\includegraphics{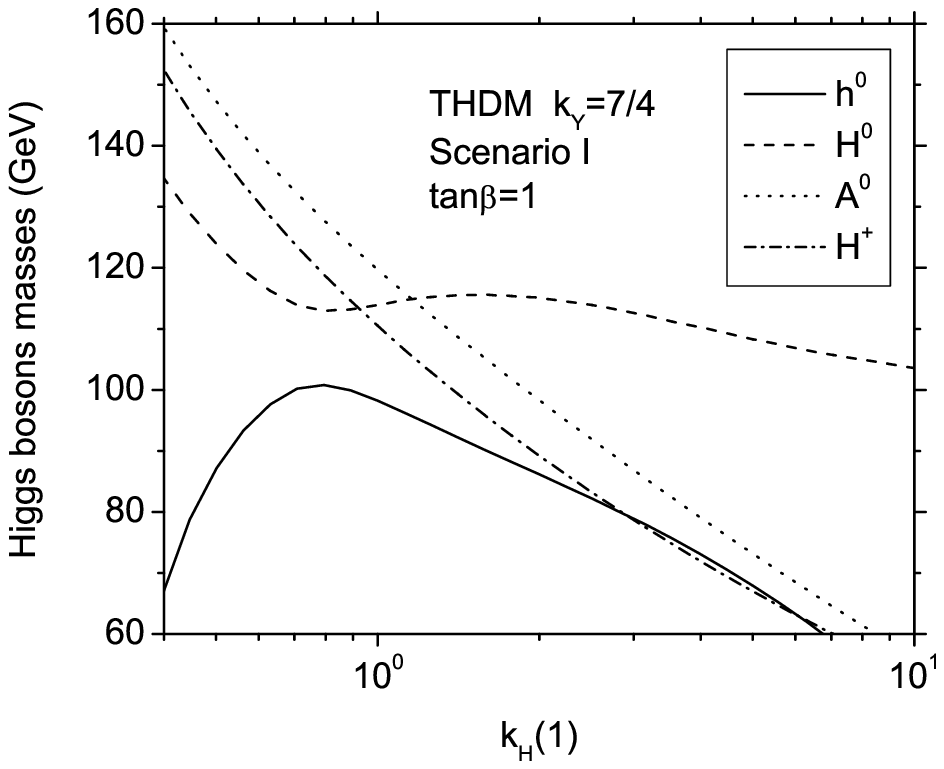}
\caption{Prediction for the Higgs bosons masses as functions of
$k_H(1)$ in the context of the THDM with $k_Y=7/4$, in the frame
of scenario I with $\tan\beta=1$, taking $m_{top}=170.0$ GeV.}
\label{figure15}
\end{center}
\end{figure}

\begin{figure}[floatfix]
\begin{center}
\includegraphics{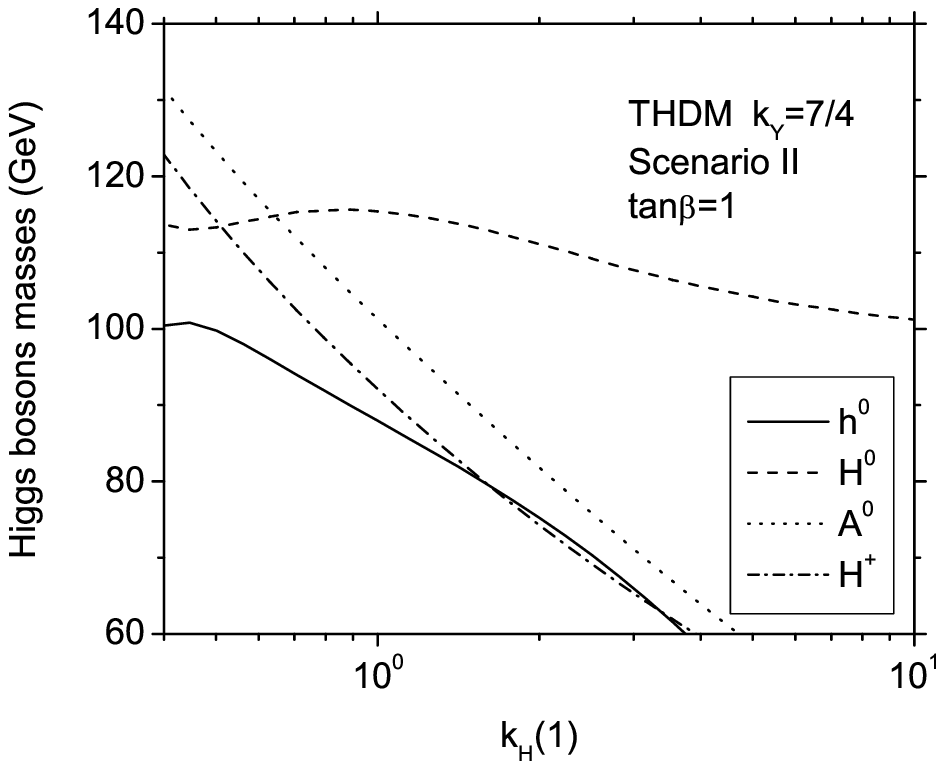}
\caption{Prediction for the Higgs bosons masses as functions of
$k_H(1)$ in the context of the THDM with $k_Y=7/4$, in the frame
of scenario II with $\tan\beta=1$, taking $m_{top}=170.0$ GeV.}
\label{figure16}
\end{center}
\end{figure}

\begin{figure}[floatfix]
\begin{center}
\includegraphics{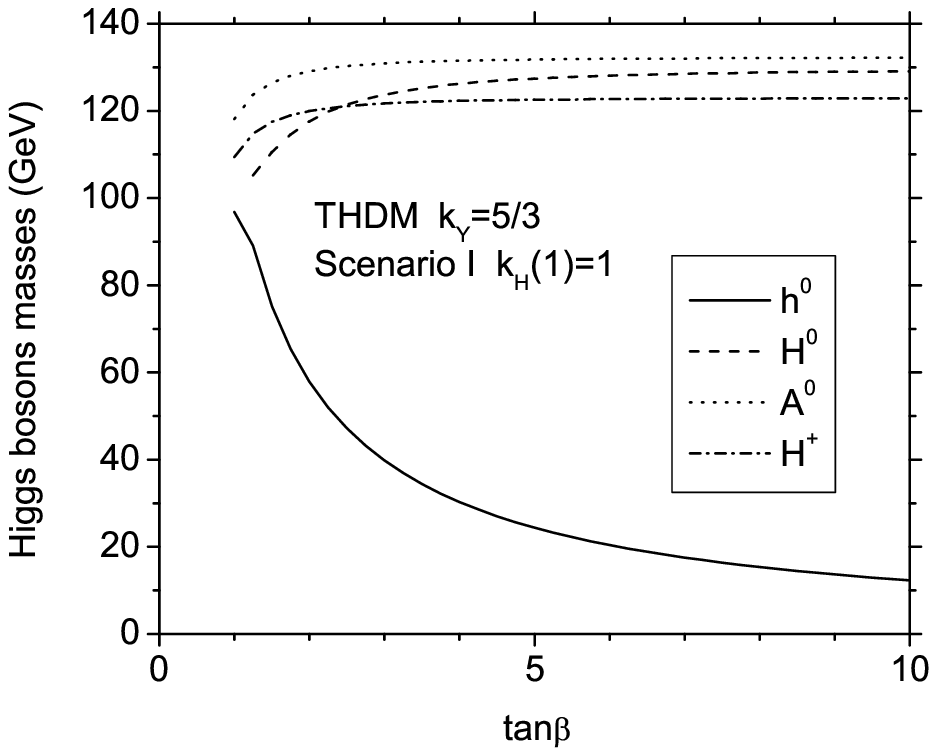}
\caption{Prediction for the Higgs bosons masses as functions of
$\tan\beta$ in the context of the THDM with $k_Y=5/3$, in the
frame of scenario I with $k_H(1)=1$, taking $m_{top}=170.0$ GeV.}
\label{figure17}
\end{center}
\end{figure}

\begin{figure}[floatfix]
\begin{center}
\includegraphics{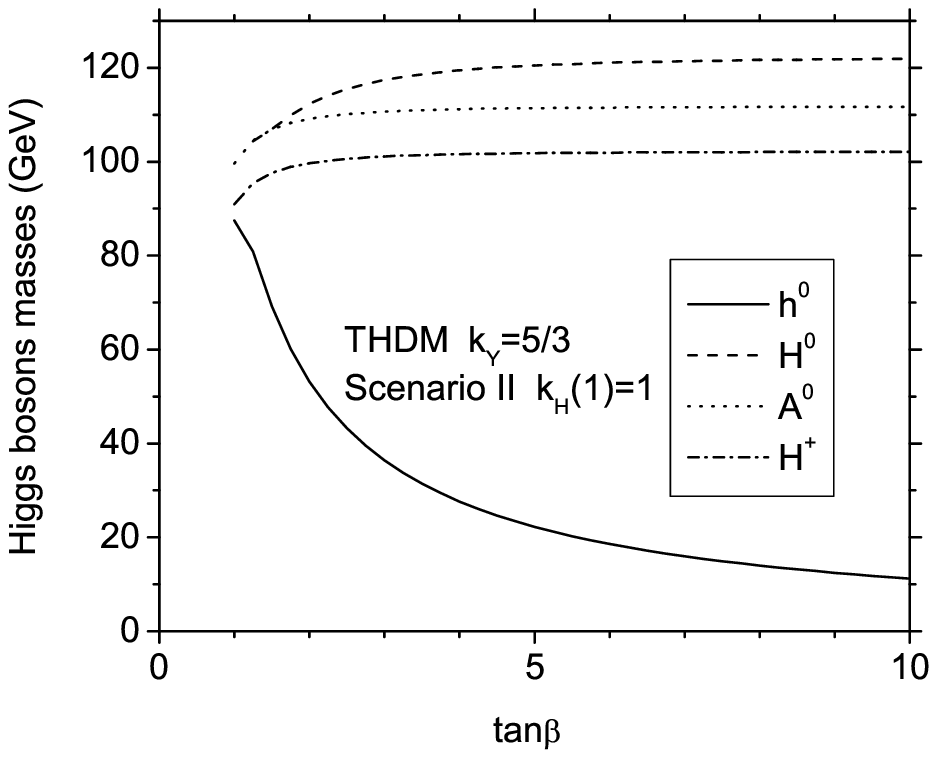}
\caption{Prediction for the Higgs bosons masses as functions of
$\tan\beta$ in the context of the THDM with $k_Y=5/3$, in the
frame of scenario II with $k_H(1)=1$, taking $m_{top}=170.0$ GeV.}
\label{figure18}
\end{center}
\end{figure}

\newpage

\begin{figure}[floatfix]
\begin{center}
\includegraphics{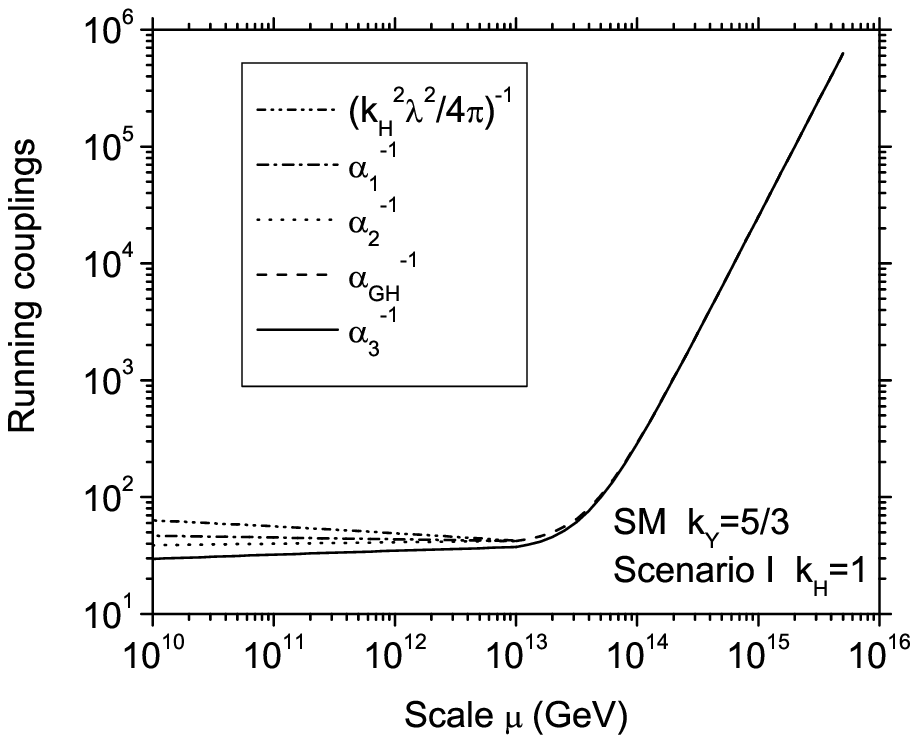}
\caption{Evolution of the running couplings as functions of the
scale $\mu$ in the context of the SM with $k_Y=5/3$, in the frame
of scenario I with $k_H(1)=1$, taking $m_{top}=170.0$ GeV.}
\label{figure19}
\end{center}
\end{figure}

\begin{table}
\begin{tabular}{|c|c|c|}
\hline

 $k_H(1)$ & $m_{h^0}$ (GeV) & $\sin^{2}(\alpha-\beta)$ \\

\hline \hline

$\;0.501\;$&$\;80.57\;$&$\;0.0227\;$\\

$\;0.562\;$&$\;87.54\;$&$\;0.0416\;$\\

$\;0.631\;$&$\;92.50\;$&$\;0.0800\;$\\

$\;0.708\;$&$\;95.78\;$&$\;0.1572\;$\\

$\;0.794\;$&$\;97.47\;$&$\;0.2919\;$\\


$\;1.000\;$&$\;96.81\;$&$\;0.6030\;$\\


$\;1.259\;$&$\;93.61\;$&$\;0.7489\;$\\


$\;1.585\;$&$\;89.86\;$&$\;0.7940\;$\\


$\;1.995\;$&$\;85.99\;$&$\;0.7985\;$\\


$\;2.512\;$&$\;81.96\;$&$\;0.7815\;$\\


\hline
\end{tabular}
\caption{Prediction for the Higgs boson $h^0$ mass and
$\sin^{2}(\alpha-\beta)$ as functions of $k_H(1)$ in the context
of the THDM with $k_Y=5/3$, in the frame of scenario I with
$\tan\beta=1$, taking $m_{top}=170.0$ GeV.}\label{tab:t1}
\end{table}

\begin{table}
\begin{tabular}{|c|c|c|}
\hline

 Scale $\mu$ (GeV) & $\alpha_3^{-1}$ & $\alpha_{GH}^{-1}$ \\

\hline \hline

$\; 1.000 \times 10^{15}\;$&$\; 0.25039991 \times 10^{5}\;$&$\; 0.25041186 \times 10^{5}\;$\\

$\; 1.259 \times 10^{15}\;$&$\; 0.39662577 \times 10^{5}\;$&$\; 0.39663589 \times 10^{5}\;$\\

$\; 1.585 \times 10^{15}\;$&$\; 0.62837664 \times 10^{5}\;$&$\; 0.62838493 \times 10^{5}\;$\\

$\; 1.995 \times 10^{15}\;$&$\; 0.99567553 \times 10^{5}\;$&$\; 0.99568198 \times 10^{5}\;$\\

$\; 2.512 \times 10^{15}\;$&$\; 0.15778035 \times 10^{6}\;$&$\; 0.15778082 \times 10^{6}\;$\\

$\; 3.162 \times 10^{15}\;$&$\; 0.25004127 \times 10^{6}\;$&$\; 0.25004155 \times 10^{6}\;$\\

$\; 3.981 \times 10^{15}\;$&$\; 0.39626483 \times 10^{6}\;$&$\; 0.39626492 \times 10^{6}\;$\\

$\; 4.467 \times 10^{15}\;$&$\; 0.49885724 \times 10^{6}\;$&$\; 0.49885724 \times 10^{6}\;$\\

$\; 5.012 \times 10^{15}\;$&$\; 0.62801339 \times 10^{6}\;$&$\; 0.62801331 \times 10^{6}\;$\\

$\; 6.310 \times 10^{15}\;$&$\; 0.99530997 \times 10^{6}\;$&$\; 0.99530970 \times 10^{6}\;$\\

$\; 7.943 \times 10^{15}\;$&$\; 0.15774357 \times 10^{7}\;$&$\; 0.15774352 \times 10^{7}\;$\\

$\; 1.000 \times 10^{16}\;$&$\; 0.25000426 \times 10^{7}\;$&$\; 0.25000419 \times 10^{7}\;$\\

\hline
\end{tabular}
\caption{Evolution of the running couplings $\alpha_3^{-1}$ and
$\alpha_{GH}^{-1}$ as functions of the scale $\mu$ in the context of
the SM with $k_Y=5/3$ (hence $M_{GH} = 1.0 \times 10^{13}$ GeV), in
the frame of scenario I with $k_H(1)=1$, taking $m_{top}=170.0$
GeV.}\label{tab:t2}
\end{table}


\begin{thebibliography}{99}

\bibitem{radcorrs1} For a review of radiative corrections see:
[LEP Collaborations],
arXiv:hep-ex/0412015.

\bibitem{radcorrs2}See also: P.~Langacker,
arXiv:hep-ph/0211065.

\bibitem{hixjenser}
U.~Baur {\it et al.}  [The Snowmass Working Group on Precision Electroweak
                  Measurements Collaboration],
in {\it Proc. of the APS/DPF/DPB Summer Study on the Future of Particle Physics
(Snowmass 2001) } ed. N.~Graf,
eConf {\bf C010630}, P1WG1 (2001)
[arXiv:hep-ph/0202001].

\bibitem{myghyunif}
  J.~L.~Diaz-Cruz,
  Mod.\ Phys.\ Lett.\ A {\bf 20}, 2397 (2005)
  [arXiv:hep-ph/0409216].

\bibitem{softsusy}
For a review see:
D.~J.~H.~Chung, L.~L.~Everett, G.~L.~Kane, S.~F.~King, J.~Lykken and L.~T.~Wang,
arXiv:hep-ph/0312378.

\bibitem{gutrev} For a recent review of GUTs see:
R.N. Mohapatra, hep-ph/0412050.

\bibitem{myradferm1}
J.~L.~Diaz-Cruz, H.~Murayama and A.~Pierce,
Phys.\ Rev.\ D{\bf 65}, 075011 (2002) [arXiv:hep-ph/0012275].

\bibitem{myradferm2}
J.~L.~Diaz-Cruz and J.~Ferrandis,
arXiv:hep-ph/0504094.

\bibitem{ADD1} N.~Arkani-Hamed, S.~Dimopoulos and G.~R.~Dvali,
Phys.\ Lett.\ B{\bf 429}, 263 (1998).

\bibitem{ADD2} I.~Antoniadis, N.~Arkani-Hamed, S.~Dimopoulos and G.~R.~Dvali,
Phys.\ Lett.\ B{\bf 436}, 257 (1998).

\bibitem{ADD3} D. Cremades, L.E. Iba\~nez and F. Marchesasno,
Nucl. Phys.\ B{\bf 643}, 93 (2002).

\bibitem{ADD4} C. Kokorelis, Nucl. Phys.\ B{\bf 677}, 115 (2004).

\bibitem{RanSundrum} L.~Randall and R.~Sundrum,
Phys.\ Rev.\ Lett.\  {\bf 83}, 3370 (1999).

\bibitem{nimashmaltz}
N.~Arkani-Hamed and M.~Schmaltz,
Phys.\ Rev.\ D{\bf 61}, 033005 (2000)
[arXiv:hep-ph/9903417].

\bibitem{abdeletal1}
R.~Barbieri, P.~Creminelli and A.~Strumia,
Nucl.\ Phys.\ B{\bf 585}, 28 (2000) [arXiv:hep-ph/0002199].

\bibitem{abdeletal2}
N.~Arkani-Hamed, S.~Dimopoulos, G.~R.~Dvali and J.~March-Russell,
Phys.\ Rev.\ D{\bf 65}, 024032 (2002) [arXiv:hep-ph/9811448].

\bibitem{abdeletal3}
R.~N.~Mohapatra and A.~Perez-Lorenzana,
Phys.\ Rev.\ D{\bf 66}, 035005 (2002) [arXiv:hep-ph/0205347].

\bibitem{abdeletal4}
N.~Cosme, J.~M.~Frere, Y.~Gouverneur, F.~S.~Ling, D.~Monderen and
V.~Van Elewyck,
Phys.\ Rev.\ D{\bf 63}, 113018 (2001)
[arXiv:hep-ph/0010192].

\bibitem{myhixXD1}
A.~Aranda, C.~Balazs and J.~L.~Diaz-Cruz,
Nucl.\ Phys.\ B{\bf 670}, 90 (2003) [arXiv:hep-ph/0212133].

\bibitem{myhixXD2}
A.~Aranda and J.~L.~Diaz-Cruz,
Mod.\ Phys.\ Lett.\ A{\bf 20}, 203 (2005)
[arXiv:hep-ph/0207059].

\bibitem{hallnomu1}
K.~R.~Dienes, E.~Dudas and T.~Gherghetta,
Phys.\ Lett.\ B{\bf 436}, 55 (1998).

\bibitem{hallnomu2}
L.~J.~Hall, Y.~Nomura and D.~R.~Smith,
Nucl.\ Phys.\ B{\bf 639}, 307 (2002)
[arXiv:hep-ph/0107331].

\bibitem{fathix} See for instance the so-called  fat Higgs
models: R. Harnik et al., Phys. Rev. D{\bf 70} (2004) 015002.

\bibitem{littlehix} For a review of little Higgs models see:
Jay G. Wacker, arXive: hep-ph/0208235.

\bibitem{hixdual} For models with composite Higgs using AdS-CFT
duality see: K. Agashe, R. Contino and A. Pomarol,
Nucl. Phys. B{\bf 719} (2005) 165.

\bibitem{XDGHix1}
N.~S.~Manton,
Nucl.\ Phys.\ B{\bf 158}, 141 (1979).

\bibitem{XDGHix2}
  D.~B.~Fairlie,
  Phys.\ Lett.\ B {\bf 82}, 97 (1979).

\bibitem{ABQuiros1}
I.~Antoniadis, K.~Benakli and M.~Quiros,
New J.\ Phys.\  {\bf 3}, 20 (2001) [arXiv:hep-th/0108005].

\bibitem{ABQuiros2}
Y.~Hosotani, S.~Noda and K.~Takenaga,
Phys.\ Lett.\ B{\bf 607}, 276 (2005) [arXiv:hep-ph/0410193].

\bibitem{ABQuiros3}
C.~Csaki, C.~Grojean and H.~Murayama,
Phys.\ Rev.\ D{\bf 67}, 085012 (2003) [arXiv:hep-ph/0210133].

\bibitem{Hosotani1}
Y.~Hosotani, S.~Noda and K.~Takenaga,
Phys.\ Lett.\ B{\bf 607}, 276 (2005) [arXiv:hep-ph/0410193].

\bibitem{Hosotani2}
C.~A.~Scrucca, M.~Serone, L.~Silvestrini and A.~Wulzer,
JHEP {\bf 0402}, 049 (2004) [arXiv:hep-th/0312267].

\bibitem{Hosotani3}
  C.~A.~Scrucca, M.~Serone and L.~Silvestrini,
  Nucl.\ Phys.\ B {\bf 669}, 128 (2003)
  [arXiv:hep-ph/0304220].

\bibitem{Hosotani4}
  G.~Burdman and Y.~Nomura,
  Nucl.\ Phys.\ B {\bf 656}, 3 (2003)
  [arXiv:hep-ph/0210257].

\bibitem{Hosotani5}
  N.~Haba and Y.~Shimizu,
  Phys.\ Rev.\ D {\bf 67}, 095001 (2003)
  [Erratum-ibid.\ D {\bf 69}, 059902 (2004)]
  [arXiv:hep-ph/0212166].

\bibitem{Hosotani6}
  I.~Gogoladze, Y.~Mimura and S.~Nandi,
  Phys.\ Lett.\ B {\bf 560}, 204 (2003)
  [arXiv:hep-ph/0301014].

\bibitem{Hosotani7}
  I.~Gogoladze, Y.~Mimura and S.~Nandi,
  Phys.\ Lett.\ B {\bf 562}, 307 (2003)
  [arXiv:hep-ph/0302176].

\bibitem{Hosotani8}
  I.~Gogoladze, Y.~Mimura and S.~Nandi,
  Phys.\ Rev.\ D {\bf 69}, 075006 (2004)
  [arXiv:hep-ph/0311127].

\bibitem{gauhixyuku1}
I.~Gogoladze, Y.~Mimura and S.~Nandi,
Phys.\ Lett.\ B{\bf 560}, 204 (2003) [arXiv:hep-ph/0301014].

\bibitem{gauhixyuku2}
I.~Gogoladze, T.~Li, Y.~Mimura and S.~Nandi,
Phys.\ Rev.\ D {\bf 72}, 055006 (2005) [arXiv:hep-ph/0504082].

\bibitem{Dienes:1996yh} K.~R.~Dienes and J.~March-Russell,
Nucl.\ Phys.\ B {\bf 479}, 113 (1996) [arXiv:hep-th/9604112].

\bibitem{rgebsm} V. Barger, M.S. Berger and M.S. Ohman,
Phys. Rev. D{\bf 47} (1993) 1093.

\bibitem{langa1} V. Barger, J. Jiang, P. Langacker and Y. Li,
Nucl. Phys. B{\bf 726} (2005) 149.

\bibitem{revPP} Review of Particle Physics. Particle data group.
Phys. Lett. B{\bf 592} (2004) 1.

\bibitem{lasttopm1}
The values for the top mass used here are based in the data
recently reported in:
  A.~Abulencia {\it et al.}  [CDF Collaboration],
   ``Measurement of the top quark mass using template methods on dilepton
  events in proton antiproton collisions at s**(1/2) = 1.96-TeV,''
  arXiv:hep-ex/0602008.

\bibitem{lasttopm2}
  See also: A.~Abulencia {\it et al.}  [CDF Collaboration],
   ``Top quark mass measurement from dilepton events at CDF II with the
  matrix-element method,''
  arXiv:hep-ex/0605118.

\bibitem{Duhrssen:2004cv1}
M.~Duhrssen, S.~Heinemeyer, H.~Logan, D.~Rainwater, G.~Weiglein and
D.~Zeppenfeld,
Phys.\ Rev.\ D {\bf 70}, 113009 (2004) [arXiv:hep-ph/0406323].

\bibitem{Duhrssen:2004cv2}
J.~L.~Diaz-Cruz and D.~A.~Lopez-Falcon,
Phys.\ Lett.\ B {\bf 568}, 245 (2003) [arXiv:hep-ph/0304212].

\bibitem{thdmky} D. Kominis and R. Sekhar, Phys. Lett. B{\bf 304} (1993) 152.

\bibitem{loren1} J.L. Diaz-Cruz and A. Mendez, Nucl. Phys. B{\bf 380},
39 (1992).

\bibitem{unknown:2006cr}
    [ALEPH Collaboration],
  arXiv:hep-ex/0602042.

\bibitem{Barate:2003sz}
  R.~Barate {\it et al.}  [LEP Working Group for Higgs boson searches],
  Phys.\ Lett.\ B {\bf 565}, 61 (2003)
  [arXiv:hep-ex/0306033].

\bibitem{futurework} J.L. Diaz-Cruz et al., work in progress.

\bibitem{splitsusy}
N.~Arkani-Hamed and S.~Dimopoulos,
arXiv:hep-th/0405159.

\bibitem{veltmanqd}
M.~J.~G.~Veltman,
Acta Phys.\ Polon.\ B{\bf 12}, 437 (1981).

\end{thebibliography}
\end{document}